\documentclass[acmtog]{acmart}

\usepackage{booktabs} 
\usepackage{xcolor}
\usepackage{float}

\usepackage{accents}
\usepackage{paralist}
\usepackage{listings}
\usepackage{sourcecodepro}
\usepackage[T1]{fontenc}
\usepackage{multicol}
\usepackage{subcaption}
\usepackage{stfloats}
\usepackage{soul}
\usepackage{multirow}
\usepackage{mathtools, nccmath}
\usepackage{graphicx}
\usepackage{bm}
\usepackage[mathscr]{euscript}
\usepackage{graphicx}  

\usepackage{epstopdf}
\usepackage[ruled]{algorithm2e}

\usepackage{microtype}

\newcommand{\todo}[1]{}
\renewcommand{\todo}[1]{{\color{red} TODO: {#1}}}

\renewcommand*\copyright{{\textcopyright}}
\usepackage{hyperref}
\hypersetup{
    colorlinks=true,
    linkcolor=blue,
    filecolor=magenta,      
    urlcolor=cyan,
}

\newcommand{\changed}[2]{#2}

\definecolor{lightblue}{RGB}{240,245,255}
\definecolor{darkblue}{RGB}{40,40,85}

\usepackage[ruled]{algorithm2e} 

\SetAlFnt{\small}
\SetAlCapFnt{\small}
\SetAlCapNameFnt{\small}
\SetAlCapHSkip{0pt}
\usepackage{enumitem}
\setitemize{noitemsep,topsep=0pt,parsep=0pt,partopsep=0pt,leftmargin=*}

\newcommand{\chongyang}[1]{\textcolor{blue}{Chongyang: #1}}

\newcommand{\simfunc}{\mathscr{F}}
\newcommand{\quantsimfunc}{\widetilde{\simfunc}}

\newcommand{\systemconfig}{\mathbf{s}}
\newcommand{\quantsystemconfig}{\widetilde{\systemconfig}}

\newcommand{\measurefunc}{\mathscr{Z}}
\newcommand{\measureval}{z}

\newcommand{\coords}{\mathbf{q}} 
\newcommand{\floatpoint}[1]{\lstinline{float#1}}

\setcopyright{acmcopyright}\acmJournal{TOG}
\acmYear{2022}\acmVolume{41}\acmNumber{4}\acmArticle{51}\acmMonth{7}
\acmDOI{10.1145/3528223.3530154}

\begin{document}

\title{Automatic Quantization for Physics-Based Simulation}

\author{Jiafeng Liu}
\email{jiafengliu@zju.edu.cn}
\affiliation{
\institution{State Key Laboratory of CAD\&CG, Zhejiang University}
\country{China}
}
\authornote{Joint first authors.}

\author{Haoyang Shi}
\email{shay@zju.edu.cn}
\affiliation{
\institution{State Key Laboratory of CAD\&CG, Zhejiang University}
\country{China}
}
\authornotemark[1]

\author{Siyuan Zhang}
\email{zhang_sy@zju.edu.cn}
\affiliation{
\institution{State Key Laboratory of CAD\&CG, Zhejiang University}
\country{China}
}

\author{Yin Yang}
\email{yin5@clemson.edu}
\affiliation{
\institution{Clemson University \& University of Utah}
\country{USA}
}

\author{Chongyang Ma}
\email{chongyangma@kuaishou.com}
\affiliation{
\institution{Kuaishou Technology}
\country{China}
}

\author{Weiwei Xu}
\email{xww@cad.zju.edu.cn}
\authornote{Corresponding author.}
\affiliation{
\institution{State Key Laboratory of CAD\&CG, Zhejiang University}
\country{China}
}


\authorsaddresses{Authors' addresses:  Jiafeng Liu, Haoyang Shi, Siyuan Zhang, Weiwei Xu, \nolinkurl{{jiafengliu, shay, zhang\_sy}@zju.edu.cn}, \nolinkurl{xww@cad.zju.edu.cn}, State Key Laboratory of CAD\&CG, Zhejiang University; Yin Yang, \nolinkurl{yin5@clemson.edu}, Clemson University \& University of Utah; Chongyang Ma, \nolinkurl{chongyangma@kuaishou.com}, Kuaishou Technology.}


%
%

%
%


\begin{abstract}
Quantization has proven effective in high-resolution and large-scale simulations, which benefit from bit-level memory saving. However, identifying a quantization scheme that meets the requirement of both precision and memory efficiency requires trial and error. In this paper, we propose a novel framework to allow users to obtain a quantization scheme by simply specifying either an error bound or a memory compression rate. Based on the error propagation theory, our method takes advantage of auto-diff to estimate the contributions of each quantization operation to the total error. We formulate the task as a constrained optimization problem, which can be efficiently solved with analytical formulas derived for the linearized objective function. Our workflow extends the Taichi compiler and introduces dithering to improve the precision of quantized simulations. We demonstrate the generality and efficiency of our method via several challenging examples of physics-based simulation, which achieves up to $2.5\times$ memory compression without noticeable degradation of visual quality in the results. Our code and data are available at
\url{https://github.com/Hanke98/AutoQuantizer}.


\end{abstract}


\begin{teaserfigure}
\includegraphics[width=\linewidth, trim=0px 0px 0px 10px,clip]{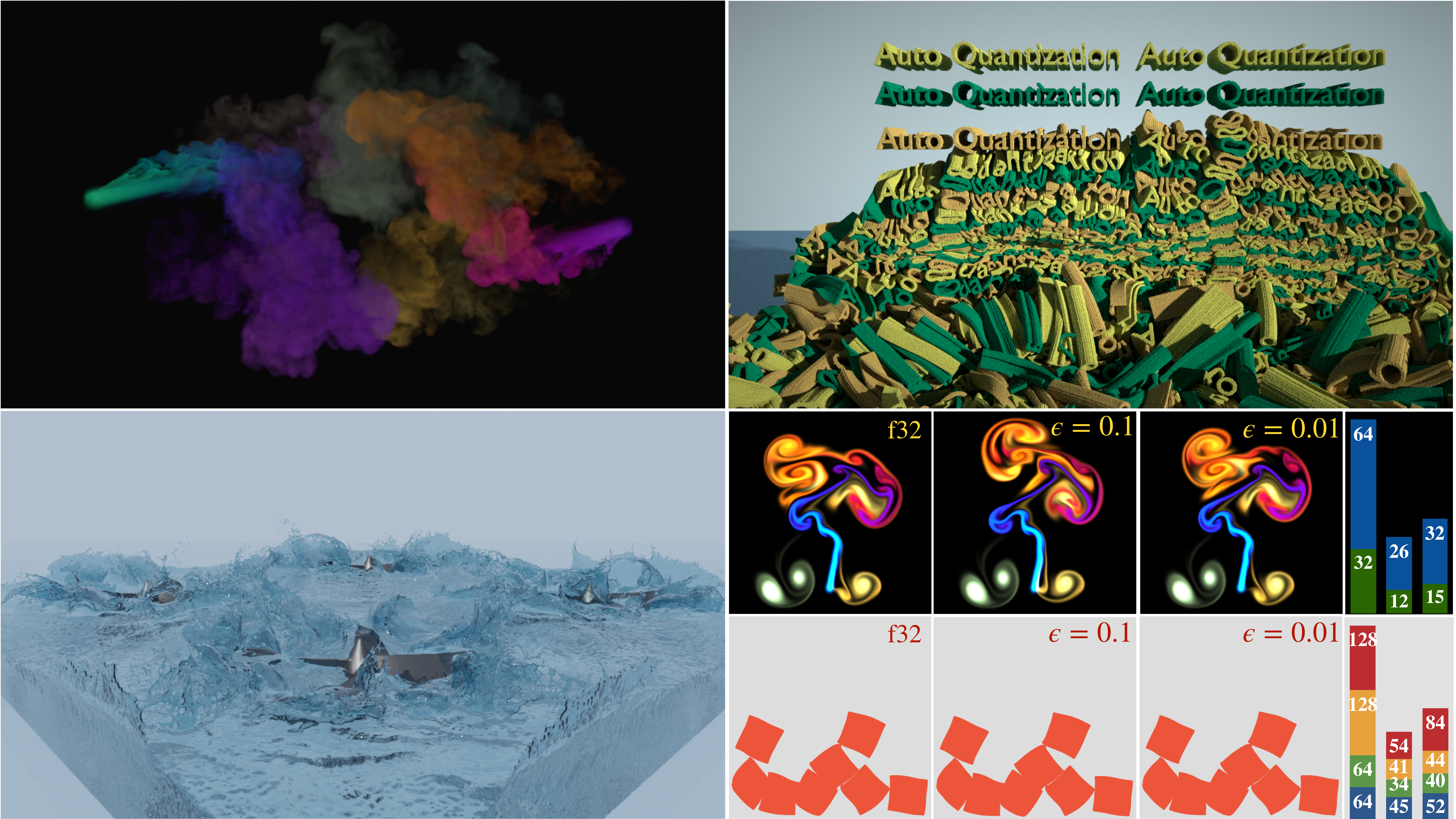}
\caption{Snapshots of our automatically quantized simulation on a single NVIDIA RTX 3090 GPU. Top left: large-scale Eulerian smoke simulation with quantized types with 230M active voxels. Top right: MLS-MPM elastics simulation of 295M particles. Bottom left: MLS-MPM fluid simulation of 400M particles. Bottom right: comparison of full-precision and quantized results in Eulerian smoke (with pressure and velocity quantized) and MLS-MPM experiment (with position, velocity, deformation gradient, and affine velocity field quantized). From left to right: \floatpoint{32} reference, fixed-point quantized result with a relative error-bound of $0.1$ and $0.01$, respectively. The total bit length is shown in the histogram on the right. Compared to the \floatpoint{32} reference, our automatically generated quantization scheme satisfies the precision requirement while achieving $2.53 \times$ and $2.04 \times$ memory compression (smoke) and $2.21 \times$ and $2.15 \times$ memory compression (elastic objects).}
\label{fig:teaser}
\end{teaserfigure}

\maketitle

\section{Introduction}
In computer graphics, the resolution of physics-based simulation plays a critical role in the visual quality of final results. Higher resolution leads to richer details and is often desirable in practice.
A relevant programming paradigm is \emph{quantization}.
For example, QuanTaichi~\cite{hu2021quantaichi} implemented a compiler-supported quantized type system, allowing the user to represent the physical variables \changed{of a simulation task}{} with fewer bits.
The simulation can then be performed in a higher resolution with the help of quantized data types.

The benefit of quantization does not come for free.
To get the best out of the quantization, users must specify an optimal quantization scheme, e.g., the fraction bit of each variable used in the simulation, to describe the data types that balance the precision and memory usages. Designing such a quantization scheme can be challenging. The problem is weaved with the following three troublesome characteristics. First, the impact of quantization on the simulation quality is not intuitive. To understand how a quantization scheme may negatively impact the simulation, one must repeat the simulation multiple times under different quantization parameters. Second, as the number of customized types increases, the search space grows exponentially. This curse of dimension makes fine-grained customization almost impossible. Third, in practice, it is easy to quantize certain properties that have a plain physical significance, e.g., positions and velocities of particles. Unfortunately, this becomes difficult when the physical significance of the property is not intuitively comprehensible, especially for non-expert users. For example, average users may find it challenging to catch the influence on the overall errors of the deformation gradient. Therefore, it is not safe to assume that quantization is a harmless gift until we find the answer to a series of questions about the impact of quantization on the simulation process. How much further can we compress the variables before the quality loss become intolerable, and how many bits of information are essential to encode the whole process? Without an automatic quantization scheme, users have to tackle these core issues manually by trial and error.

In this paper, we study how to solve the automatic quantization problem using a constrained optimization algorithm.
Our method effectively obtains a quantization scheme automatically and efficiently. Based on the uncertainty propagation theory~\cite{cohen1998introduction}, we can consider the overall error of the simulation as a synthesis of the errors induced by quantization at each step. It allows us to specify a measurement to indicate the simulation error after quantization, e.g., the volume change for an incompressible fluid simulation, and use the gradients of the simulation error concerning each quantized variable to characterize the contribution of each quantization operation. Consequently, the automatic quantization problem can be formulated as a constrained optimization problem via local linearization of the overall simulation error. By solving the optimization problem, we can achieve a desirable compression rate of memory consumption with visually plausible simulation results.

The optimal fraction bits in a quantization scheme should balance the memory consumption and the simulation precision.
However, as the number of possible fraction bits is discrete, the formulated optimization problem is computationally expensive since, in combinatorial optimization algorithms, we need to repeat the physics-based simulation multiple times to evaluate the searched solution for fraction bits.
Therefore, we convert the discrete fraction bit variable to the continuous quantization resolution variable such that the optimization problem can be solved continuously via the Lagrange multiplier method.
Our method computes gradient information after running the physical simulation once and then leverages it to obtain the fraction bits for each variable in an analytical way, providing fast user feedback.  


Although the uncertainty propagation theory can also consider the correlation among the random variables, it is expensive to evaluate the correlation among variables in a large-scale simulation problem. Therefore, we introduce \emph{dithering}~\cite{schuchman1964dither} to enforce the independence among the round-off errors so that the correlation terms in uncertainty propagation can be safely ignored. Moreover, our experiments show that dithering significantly improves simulation results in some cases since it can alleviate dependence between round-off errors in the forward simulation.

In our formulation, gradients are essential to error estimation. Unfortunately, the memory consumption of gradients computation is too large for a long-multi-step simulation because the states must be checkpointed at each time step. We \changed{take advantage of}{adopt} the methods from~\cite{braconnier2002rounding} with $\mathrm{O}(\log N)$ space complexity to save memory during the back-propagation process.
After we obtain the quantization scheme by solving the optimization, it is also problematic to automatically fit those custom data types with different bit widths into memory. The compiler has to pack those custom data into a physical word, e.g., 32-bit or 64-bit length. To tackle this problem, we extend the Taichi compiler~\cite{hu2019taichi} and offer a new bit-level container called \emph{bit pack} that allows users to place arbitrary lengths of custom data types into memory regardless of the limits of physical word.

In summary, our main contributions include:
\begin{enumerate}
    \item A novel formulation and solution for computing a quantization scheme based on uncertainty propagation theory;
    \item A simple compiler interface for \changed{variable}{} quantization and a low memory cost implementation of automatic differentiation;
    \item Comprehensive evaluation and validation of the proposed automatic quantization scheme with empirical instructions on how to use our method.
\end{enumerate}

\begin{figure*}
    \centering
    \includegraphics[width=\textwidth]{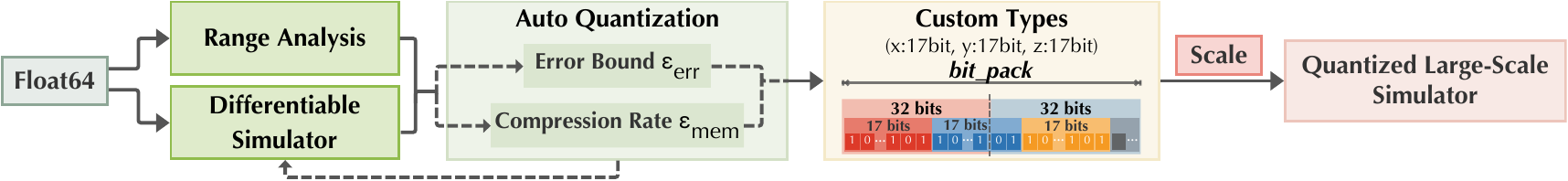}
  \caption{Overview of our method. A differentiable simulator is developed on a relatively low resolution in a generic workflow. The user specifies an evaluation function and launches the simulation with \floatpoint{64}. Ranges and derivatives are recorded during this run. Our auto-quantization system can then derive a quantization scheme according to the user's specification on either error or memory constraints. The user starts to refine the simulation on a higher resolution with the derived quantization scheme before eventually getting a quantized large-scale simulator.}
  \label{fig:flow_chart}
\end{figure*}

Our method is implemented as an extension to the Taichi programming language~\cite{hu2019taichi} to take advantage of its feature of automatic differentiation.
\changed{
However, our automatic quantization scheme is general for physics-based simulation and can be implemented on any differentiable programming systems, e.g., JAX and TensorFlow.}{However, other automatic differentiable
programming frameworks,
such as Jax\footnote{\url{https://github.com/google/jax}} and Tensorflow\footnote{\url{https://www.tensorflow.org/}},
can also be adopted to implement differentiable simulations.
Therefore, our method can be integrated into these frameworks to obtain quantization schemes as well.}
See Fig.~\ref{fig:flow_chart} for an overview of our system.

\section{Related Work}

\paragraph{High-resolution simulation}
QuanTaichi \cite{hu2021quantaichi} introduces quantized data types into the area of physics-based simulations.
Empowered by bit-level compression, rich details can be achieved in the simulation results. Besides, sparse data structures \cite{setaluri2014spgrid, wu2018fast, museth2013vdb, hu2019taichi} are also important to implement high-resolution simulations. These techniques aim to allocate memory only when necessary.

However, if all memory is occupied, it is more sensible to use different resolutions/scales in the simulation scenes \cite{solenthaler2011two, wang2018parallel}.
It is also shown that great visual details can be revealed by adapting the resolution dynamically \cite{aanjaneya2017power, ferstl2014large, de2015power}. Coarsening the interior elements is a commonly used method. Besides, the Boundary Element Method (BEM)~\cite{james1999artdefo} is another practical way to enhance the details of simulation \cite{keeler2014ocean, hahn2015high, huang2021ships}. For all methods above, parallelization on multi-GPUs is essential to increasing the upper bound of the simulation scale \cite{gao2018gpu, wang2020massively, wu2015system,liu2016scalable}.
Most of these methods need tailored implementation of the corresponding algorithms, while our method allows users to reduce memory consumption easily with minor modification of the original simulator. 

\paragraph{Word-length optimization}
\changed{}{In Digital Signal Processing (DSP), word-length optimization (WLO)~\cite{catthoor1988simulated, sung1995simulation, Constantinides2001} is a methodology to determine the quantization schemes of data to trade-off between precision and computational cost on resource-limited devices.} The key component of \changed{determine a set of good data representations}{WLO} is an evaluation method that approximates the relationship between the quantization errors and the system precision.
Early research uses simulation-based methods \cite{sung1995simulation} to evaluate a set of quantization schemes, which becomes intolerable as the scale and number of variables increase. \cite{Constantinides2001} uses transfer functions of linear time-invariant (LTI) systems to estimate the system errors.
Perturbation analysis is introduced to obtain the error model for non-linear systems, in which the main idea is to consider the quantization error as a small perturbation to the signal and to measure the contribution of each quantization operation to the systematic error.
\cite{shi2004perturbation} fits an error model by reducing the word length and catching the variation of the total error.
\cite{constantinides2006word} performs a Taylor expansion and uses the first-order derivatives as the coefficients of the linear combination.
Another method of constructing the error model is Interval Arithmetic (IA) \cite{moore1996interval}. IA is a widely-used method for precision analysis. In IA, numbers are represented not only by themselves but also by a lower and an upper bound. An entire set of arithmetic rules is defined to support the range and error analysis.
Typical improvement on IA is Affine Arithmetics (AA), which takes the correlations of signals into consideration~\cite{lee2006accuracy, vakili2013enhanced}.

\changed{The formulation of world-length optimization has been proposed in \cite{Constantinides2001} and has evolved in follow-up research.}{} The main form of WLO is to minimize the hardware cost of the whole system under constraints of the total errors. Because the quantization schemes are represented by integers, solving the optimization is not an easy task.
There are two practical ways to solve the optimization.
One is to adopt stochastic search methods such as simulated annealing \cite{constantinides2003perturbation}.
The other methodology is setting the fraction bits to the maximum (e.g., 32) at first and then gradually reducing them to find the minimal feasible word length \cite{shi2004perturbation}.

The intrinsic nature of these DSP tasks indicates that they can yield reasonable results within a short period of time. However, for high-resolution physics-based simulation, fitting the error model by running the simulation multiple times or solving the optimizations by a stochastic search method can be time-consuming.

\paragraph{Differentiable simulations}
Running a differentiable simulation is an upstream task for our core algorithm of automatic quantization. \changed{}{Recently, several types of differentiable-simulation methods have been developed. ChainQueen~\cite{hu2019chainqueen} presents a differentiable soft body simulator using the Material Point Method (MPM) with explicit time integration.
A differentiable finite element method (FEM) with implicit time integration has also been proposed under the framework of projective dynamics~\cite{du2021diffpd,bouaziz2014projective}.
Another line of methods approximates the simulation process via neural networks~\cite{battaglia2016interaction, li2018learning, sanchez2020learning}, which inherits the differentiability from the network. However, the differentiation is not rigorously built on physics, and the physical correctness might not be guaranteed in this way~\cite{du2021diffpd,bangaru2021systematically}.
In addition, Automatic differentiation tools are used to reduce the implementation efforts of differentiable simulations~\cite{de2018end, degrave2019differentiable}.
For instance,  DiffTaichi~\cite{hu2019difftaichi} proposes a programming framework that automatically generates gradient computation kernels that can be applied to implement various simulation algorithms. As for optimal control tasks, the adjoint method is often utilized to compute the gradients of a loss function with respect to simulation parameters~\cite{geilinger2020add,mcnamara2004fluid}. To simplify the programming process and facilitate integration with the compiler, we choose to extend the DiffTaichi framework in our implementation.}

Prior to our work, applications of differentiable simulation include parameter estimation, inverse dynamics, and motion control~\cite{liang2019differentiable,qiao2020scalable}.
Our work identifies its significance as the sensitivity of the target function to errors in each input variable and utilizes it as the key ingredient in error estimation.

\paragraph{Dithering}
In our framework, we leverage dithering as a method to modulate the distribution of quantization errors.
Dithering refers to imposing a random noise on the continuous signal before quantization. It has long been applied in improving the subjective quality of quantized representation for images or speeches \cite{waveform}.
Schuchman~\shortcite{schuchman1964dither} develops the theory about the property of dithering and proves that a desirable dither signal should generate quantization errors independent of the input signal.
The common applications use the manner of subtractive dithering, which enjoys strong mathematical robustness, and is considered to be an efficient algorithm to send a real number using only one bit.
Subtractive dithering assumes that the receiver is informed of the random noise, which is subtracted out when reproducing the original signal.
However, difficulty in fulfilling this assumption has led to research on non-subtractive dithering with similar effects on the moments of quantization error \cite{nonsubtractive}.
We follow the non-subtractive approach in our implementation since bookkeeping the dithering signal violates our motivation to reduce memory consumption in the first place.

\section{Background}

 \subsection{Quantization Procedure for Physics-Based Simulation}
Our algorithm returns a quantization scheme of a simulation task, given a memory budget or an accuracy expectation specified by the user.
Since a physical simulation proceeds one time step after another, this process is inherently sequential.
The error induced by quantized computation accumulates along the simulation procedure.
\changed{}{Table~\ref{tab:notations} summarizes the notations used in our algorithm for clarity.}

\begin{table}
\centering
\caption{\changed{}{List of notations.}}
\begin{tabular}{l|l|l}
\toprule
\textbf{Symbol}     & \textbf{Type}     & \textbf{Meaning or Definition}                          \\ \midrule \midrule
$\simfunc$          & function & simulation function of one time step           \\
$\systemconfig$     & vector   & system state                                   \\
$\quantsimfunc$     & function & quantized simulation function  \\
$\quantsystemconfig$& vector   & quantized system state                         \\
$\coords$           & vector   & component of $\systemconfig$             \\
$\mathbf{e}$        & vector   & error vector                                   \\
$\measurefunc$      & function & evaluation function                            \\
$z$                 & scalar   & value of the evaluation function            \\
$x$                 & scalar   & arbitrary variable with quantized type             \\
$\triangle$         & scalar   & quantization resolution                        \\
$b$                 & scalar   & fraction bits for fixed-point numbers                 \\
$R$                 & scalar   & range of the fixed-point numbers      \\
$H$                 & scalar   & number of quantized data types   \\
$P$                 & scalar   & number of variables          \\
$T$                 & scalar   & total simulation steps          \\
$N$                 & scalar   & total quantization times   \\
$\phi$              & function & quantization encoding function                 \\
$\phi^{-1}$         & function & quantization dencoding function                 \\
$\sigma$            & scalar   & standard deviation                            \\
$\epsilon_{err}$    & scalar   & target simulation error tolerance              \\
$\epsilon_{mem}$    & scalar   & target memory compression rate                 \\
$M$         & scalar   & reference memory consumption                  \\
$\mathscr{P}$      & function & probability density function    \\
$\mathscr{G}$      & function & adjoint kernel for computing gradients    \\
$\mathbf{g}$          & vector & accumulated squared gradients    \\
\bottomrule
\end{tabular}
\label{tab:notations}
\end{table}

\begin{figure}
\centering
\includegraphics[width=\linewidth]{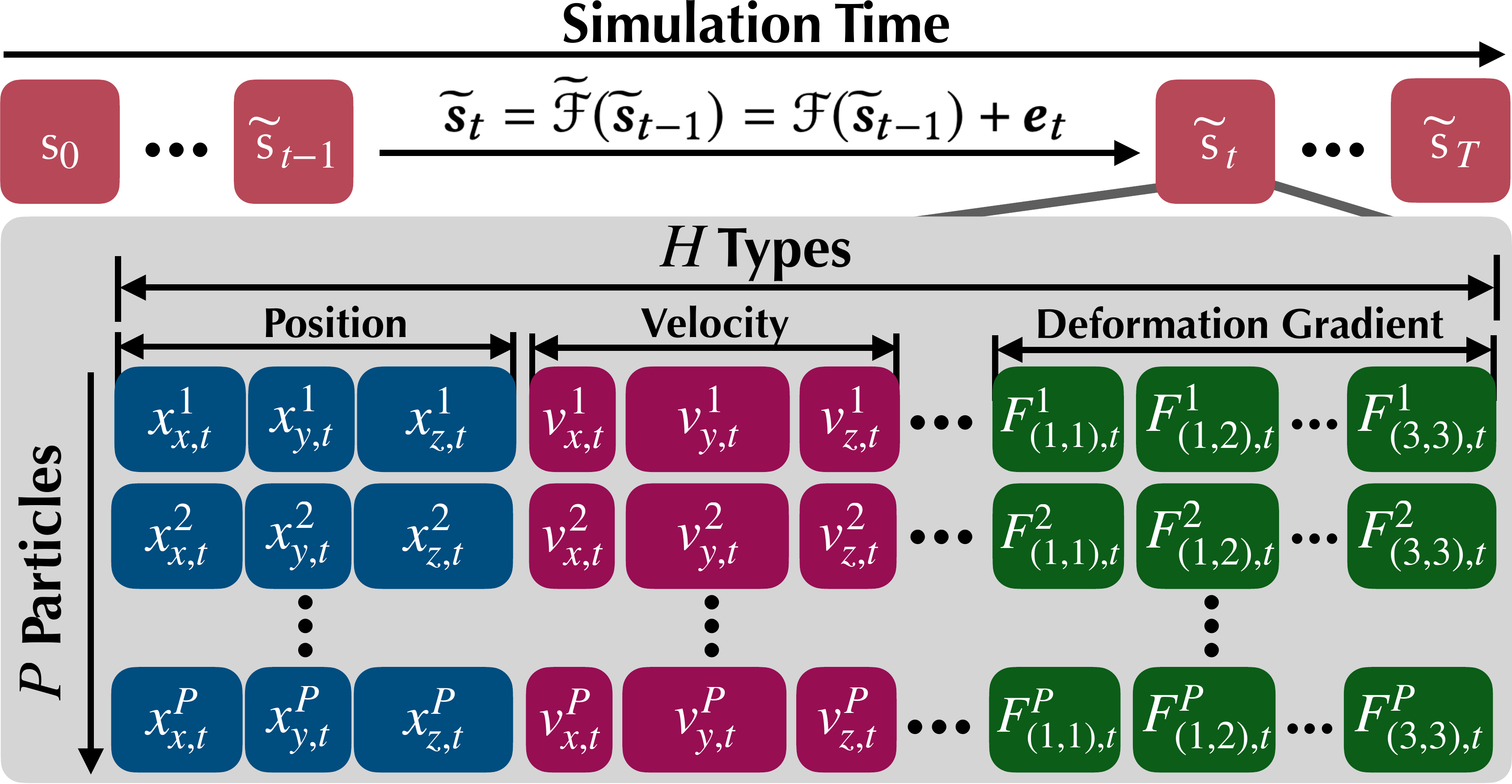} 
\caption{Notations of a quantized MPM simulation. In this case, we only quantize the attributes stored on particles, so there are $P$ variables for each attribute, where $P$ is the number of particles.}
\label{fig:notations}
\end{figure}

Let $\simfunc$ denote a generic simulation function of one time step, which maps the system configuration from the previous time step $\systemconfig_{t-1}$ to the current step $\systemconfig_t$, i.e., $\systemconfig_t = \simfunc(\systemconfig_{t-1})$. Meanwhile, we have a quantized simulation function, $\quantsimfunc$, which limits the total bit length of $\systemconfig_t$ and results in a quantized simulation result $\quantsystemconfig_t$. $\quantsimfunc$ can also be understood as imposing a quantization error $\mathbf{e}_t$ to the full-precision simulation $\systemconfig_t$ by eliminating nonzero quantities at its least significant digits. If $\systemconfig_{t-1}$ is quantized, we have:
\begin{align}
    \quantsystemconfig_t = \quantsimfunc(\quantsystemconfig_{t-1}) = \simfunc(\quantsystemconfig_{t-1}) + \mathbf{e}_t.
\end{align}
During the simulation, $\systemconfig$ is composed of several generalized coordinates and other attributes, e.g., particle positions, velocities etc: $\systemconfig = (\coords_1^T, \coords_2^T, \cdots, \coords_H^T)^T$, where $H$ is the number of different physical quantities, such as components of position and velocity involved in the simulation, and each generalized coordinate $\coords_i$ is of $P_i$ dimension. For example, in material point method (MPM), $P_i$ can be the number of particles or grid cells (see Fig.~\ref{fig:notations}).

We assume that the user appoints a specific evaluation function, $\measurefunc$, which measures a metric of interest of the simulation over $T$ time steps. Naturally, the goal of an automatic quantizer seeks to minimize $\delta \measureval$, the difference between the evaluation function values of the quantized and full-precision simulations:
\begin{align}
    \delta \measureval = \left|\measurefunc\big(\quantsystemconfig_0,\quantsystemconfig_1,\cdots,\quantsystemconfig_T\big) - \measurefunc\big(\systemconfig_0,\systemconfig_1,\cdots,\systemconfig_T\big) \right|^2.
\label{eq:detla_measure}
\end{align}
For example, $\measurefunc$ can be set to calculate the kinetic energy of all the particles in a fluid simulation, and then $\delta z$ should identify the magnitude of the numerical viscosity induced by the quantization.

We follow the same quantization algorithm as in~\cite{hu2021quantaichi}. A quantized type variable is stored in the memory with a limited digit count but decoded as a standard \floatpoint{32} or \floatpoint{64} quantity when participating in the computation. Typically, this operation is lossless.
The reverse operation i.e., the encoding from a regular float to the quantized type, however, brings a quantization error.

Consider a quantization operation of a fixed-point transformation $\phi : \mathbb{R} \mapsto \mathbb{Z}$, which takes a scalar $v \in \mathbb{R}$ as the input and produces its integer representation $u \in \mathbb{Z}$ and its inverse $\phi^{-1}$, estimating the floating-point value from a quantized integer $u$:
\begin{align}
    u = \phi(v) = \lceil v / \triangle \rfloor, \quad \phi^{-1}(u) = u \cdot \triangle,
\end{align}
where $\lceil \cdot \rfloor$ is the function of rounding to the nearest integer and $\triangle = 2^{-b}R$ is the resolution of the corresponding quantization, where $b$ and $R$ are the fraction bits and the range for $v$ respectively. The error induced by $h$ then becomes $e = v - \phi^{-1}\big(\phi(v)\big)$.

\changed{In this quantization scheme, we need to figure out the optimal value for $b$ and $R$, i.e., the fraction bits and the range for each custom type, to minimize the total bit length.}{In this quantized simulation procedure, we need to determine the optimal value for fraction bits $b$ and ranges $R$ for each custom type. For example, in a 3D Eulerian fluid simulation, each cell is associated with a vector of velocity ($\mathbf{v}$) and a scalar of pressure ($p$). If we want to take advantage of quantized simulation, we need to figure out how many fraction bits we should assign to each quantity and what is the feasible range. However, there are many choices for the fraction bits, i.e., 16 bits for $p$ and each component of $\mathbf{v}$ or 10 bits for $p$ and $3\times 18$ bits for $\mathbf{v}$. Different choices will lead to different simulation errors and memory compression rates. In this paper, we focus on how to automatically obtain the number of fraction bits according to users' requirements on memory budget or computation precision. As for the range analysis, we record the ranges from the full precision simulation and multiply the ranges by a factor (e.g., 2) to prevent overflowing.}

\subsection{Uncertainty Propagation Theory}
Considering the quantization operation as imposing an error $e_i$ to a variable $x_i$,  we can model the relationship between the quantity $\delta \measureval$ and system configuration variables using \emph{uncertainty propagation theory}~\cite{cohen1998introduction}. \changed{}{This theory proposes to measure the uncertainty of a variable as standard deviation and construct a step-by-step propagation procedure to estimate the final uncertainty of a composite function. Specifically, if all the input variables are uncorrelated, the uncertainty of a synthesized variable $z$ can be estimated from the uncertainty of all its contributing variables $\{ x_i \}$ } as follows:
\begin{align}\label{equ:error_prop}
    \sigma_z^2 = \sum_{i=1} \left(\frac{\partial z}{\partial x_i}\right)^2 \sigma_{x_i}^2,
\end{align}
where $\sigma_z^2$ and $\sigma_{x_i}^2$ are the variances of $z$ and $x_i$, respectively.

In our case, Eq.~\eqref{equ:error_prop} allows us to track the uncertainty from all the quantization errors for $\delta \measureval$ in Eq.~\eqref{eq:detla_measure}:
\begin{equation}
\label{equ:second_momentum}
    E[\delta z] = \sum_{i=1}^{N} \int_{-\triangle_i/2}^{\triangle_i/2} \left(\frac{\partial \measurefunc}{\partial e_i}\right)^2 e_i^2 \mathscr{P}(e_i) \mathrm{d}e_i
\end{equation}
where the summation traverses all the components in each generalized coordinate across $T$ time steps, and thus $N = \sum_{i=1}^H P_i \cdot T$.
The integral range is the absolute error bound, which is half of the quantization resolution.
The probability density function for an error component $e_i$ is denoted as $\mathscr{P}(e_i)$.

\changed{Note that our model has neglected the correlation between contributing random variables by assuming independence, hence the absence of the cross terms in Eq.~\eqref{equ:error_prop}.\chongyang{Need to rephrase.}}{Note that we assume that the quantization errors are uncorrelated, and therefore the cross terms of correlations are neglected in Eq.~\eqref{equ:error_prop}} \changed{Therefore, it}{Under this assumption, our model} lacks the ability to recognize the correlation between quantization errors. It is essentially a localized linear expansion at the point of the true value. Hence, we introduce \emph{dithering} operations so that this form of uncertainty propagation can be applied to our problem properly (see Sec.~\ref{sec:dithering}).


In our implementation, a full-precision simulation needs to be carried out first to obtain $z = \mathscr{Z}(\bm{s}_i)$, the partial derivatives of the quantization error, as well as the range of each custom data type.
\changed{To prevent overflowing, we set the range of each variable as $R_i  \leftarrow 2 \sup(|x_i|)$.}{}
\section{Uncertainty-Propagation-based Quantization}
In this section, we first describe the details of the objective functions and then present how to convert the automatic quantization problem into a continuous optimization problem using vector variables consisting of the quantization resolution $\triangle$ of each quantity.

\subsection{Formulation of Objective Functions}
The proposed automatic quantization includes two different strategies, aiming to provide the user with either a controllable accuracy bound or a guaranteed memory compression rate.
Below we elaborate the objective functions for these two strategies in detail.


\paragraph{Error-bounded quantization.}
\label{sec:error_bound_quantization}
Being quantized, simulation variables inevitably lose precision under capped bit length, which potentially impacts the simulation quality and may even result in visible artifacts.
Error-bounded quantization scheme aims to address this concern. It offers the accuracy guarantee prescribed by a user-specified error bound and lowers the total memory consumption:
\begin{equation}\label{eq:sim_err}
    \min_{b_h} \sum_{h = 1}^H P_h b_h,\quad \text{s.t.} \; E[\delta z] < (\epsilon_{err} \cdot z)^2,
\end{equation}
where $P_i b_i$ gives the memory footprint of each generalized coordinate $\coords_i$, $E[\cdot]$ computes the expectation of a random variable, $z = \mathscr{Z}\big(\systemconfig_0,\systemconfig_1,\cdots,\systemconfig_T\big)$ is the reference simulation metric obtained without quantization, and $\epsilon_{err}$ is the relative error tolerance specified by users.


\paragraph{Memory-bounded quantization.}
In a different scenario where the hardware resource is limited, a guaranteed memory consumption could be more favored in practice. To this end, we pose the memory compression rate as a hard constraint that must be fulfilled by the automatic quantization algorithm. Based on this constraint, we push to suppress the accuracy loss as much as possible: 
\begin{equation}\label{eq:sim_mem}
    \min E[\delta z] \quad \text{s.t.} \sum_{h = 1}^H P_h b_h < \epsilon_{mem} \cdot M,
\end{equation}
where $\epsilon_{mem}$ is the memory compression rate specified by the user, $M$ stands for the total memory usage for the full-precision simulation, e.g., $M = \textsf{sizeof}(\floatpoint{32})\sum P_h$, assuming \floatpoint{32} is used as the data type.

\subsection{Continuous Optimization Problem}
We change the fraction bit variable $b_h$ into its corresponding quantization resolution variable $\triangle_h$ for objective functions defined in Sec.~\ref{sec:error_bound_quantization}, such that the automatic quantization problem can be solved in a continuous way.
Assuming $\mathscr{P}(e_i)$ follows a uniform distribution, Eq.~\eqref{equ:second_momentum} is integrated into:
\begin{align}
    E[\delta z] = \frac{1}{12}\sum_{h=1}^{H}\triangle_h^2 g_h \overset{\operatorname{def}}{=} \frac{1}{12}\sum_{h=1}^{H} \triangle_h^2\left(\sum_{t = 0}^{T}{\sum_{i=1}^{P_h} \left(\frac{\partial z}{\partial x_{t,h}^i}\right)^2}\right), 
\label{eq:expectation_with_delta_h}
\end{align}
where the coefficient $g_h$ is the accumulated squared gradient of the variable with type $h$. More, specifically, since $b_h=-\log_2{\frac{\triangle_h}{R_h}}$, the objective function in Eq.~\eqref{eq:sim_err} after the variable quantization is as follows:
\begin{equation}
\begin{aligned}
    \min_{\triangle_h}\sum_{i=1}^{H}{-P_h \log_2{\frac{\triangle_h}{R_h}}}    \\
    s.t.
   \frac{1}{12}\sum_{h=1}^{H} {\triangle_h^2 g_h} < (\measureval    \epsilon_{err})^2     \\
    \triangle_h \in \mathbb{R^+}
    \label{eq:eror_bound_2}
\end{aligned}
\end{equation}
Such a continuous optimization problem has an analytical solution using Lagrange multiplier method:
\begin{align}
    \triangle_h = \sqrt{\frac{12P_h(\epsilon_{err} \cdot z)^2}{g_h \sum_{h=1}^{H}{P_h}}}
\end{align}
\changed{where $g_h = \sum_{t = 0}^{T}{\sum_{i=1}^{P_h} \left(\frac{\partial z}{\partial x_{t,h}^i}\right)^2}$.}{} The fraction bits for each type of variable are then obtained according to the value of $\triangle_h$.
Likewise, we can obtain the analytical solution for memory-bounded scenario in Eq.~\eqref{eq:sim_mem} after the variable quantization.
Please refer to section 4 in supplementary material for the solution derivations.

\paragraph{Discussions.}
The objective function in Eq.~\eqref{eq:eror_bound_2} is a local linearization of Eq.~\eqref{eq:sim_err}. Indeed, the linearization can be iteratively updated at new values of $\triangle_h$ after we solve Eq.~\eqref{eq:eror_bound_2}.
However, such an iterative algorithm incurs a high computational cost since the gradients must be computed after another round of forwarding simulation in auto-diff.
In practice, we have found linearizing using the gradients computed with full-precision simulation is able to produce approximate solutions that have the desirable properties: the quantized simulation result is visually consistent with the full-precision simulation, and their differences from the full-precision simulation or the memory footprint can be controlled in most cases using our objective functions with dithering.
Therefore, we accept the solution without iteration, which means we prefer speed over optimality in automatic quantization. It can provide users with faster feedback.
\changed{}{The pseudocode of solving the quantization scheme under error-bound constraints can be found in Algorithm~\ref{algorithm:error_bound}.}

\begin{algorithm}[t]
    \caption{\changed{}{Automatic quantization scheme based on an error-bound constraint.}}
    \label{algorithm:error_bound}
    \LinesNumbered
    \KwIn {$\epsilon_{err}$}
    \KwOut {$(b_1, R_1),...,(b_H, R_H)$}
    $T\leftarrow$ {Total simulation steps}\\
    $H\leftarrow$ {Number of quantized data types} \\
    \For{h = 1 to H}{
        $P_h \leftarrow $ {number of variables using type $h$} \
    }
    $\systemconfig_0 \leftarrow$ {Initial conditions}\\
    \For{t = 0 to T-1}{
        \tcp{step forward} \
        $\systemconfig_{t+1} \leftarrow \simfunc(\systemconfig_{t})$ \\
        \tcp{update ranges for each quantity} \
        {Update $R_1...R_H$ according to $\systemconfig_{t+1}$}
    }
    \tcp{compute the evaluation function} \
    $z \leftarrow \measurefunc\big(\systemconfig_0,\systemconfig_1,\cdots,\systemconfig_T\big)$ \\
    Back propagation and accumulate gradients in $\textbf{g}$ according to Eq.~\eqref{eq:expectation_with_delta_h} \\
    \For{h = 1 to H}{
        $\triangle_h \leftarrow \sqrt{\frac{12P_h(\epsilon_{err} \cdot z)^2}{g_h \sum_{h=1}^{H}{P_h}}}$ \\
        $b_h \leftarrow \lceil -\log_2\frac{\triangle_h}{R_h} \rceil$
    }
\end{algorithm}

\section{Dithering}
\label{sec:dithering}
\begin{figure}
    \centering
    \includegraphics[width=\linewidth]{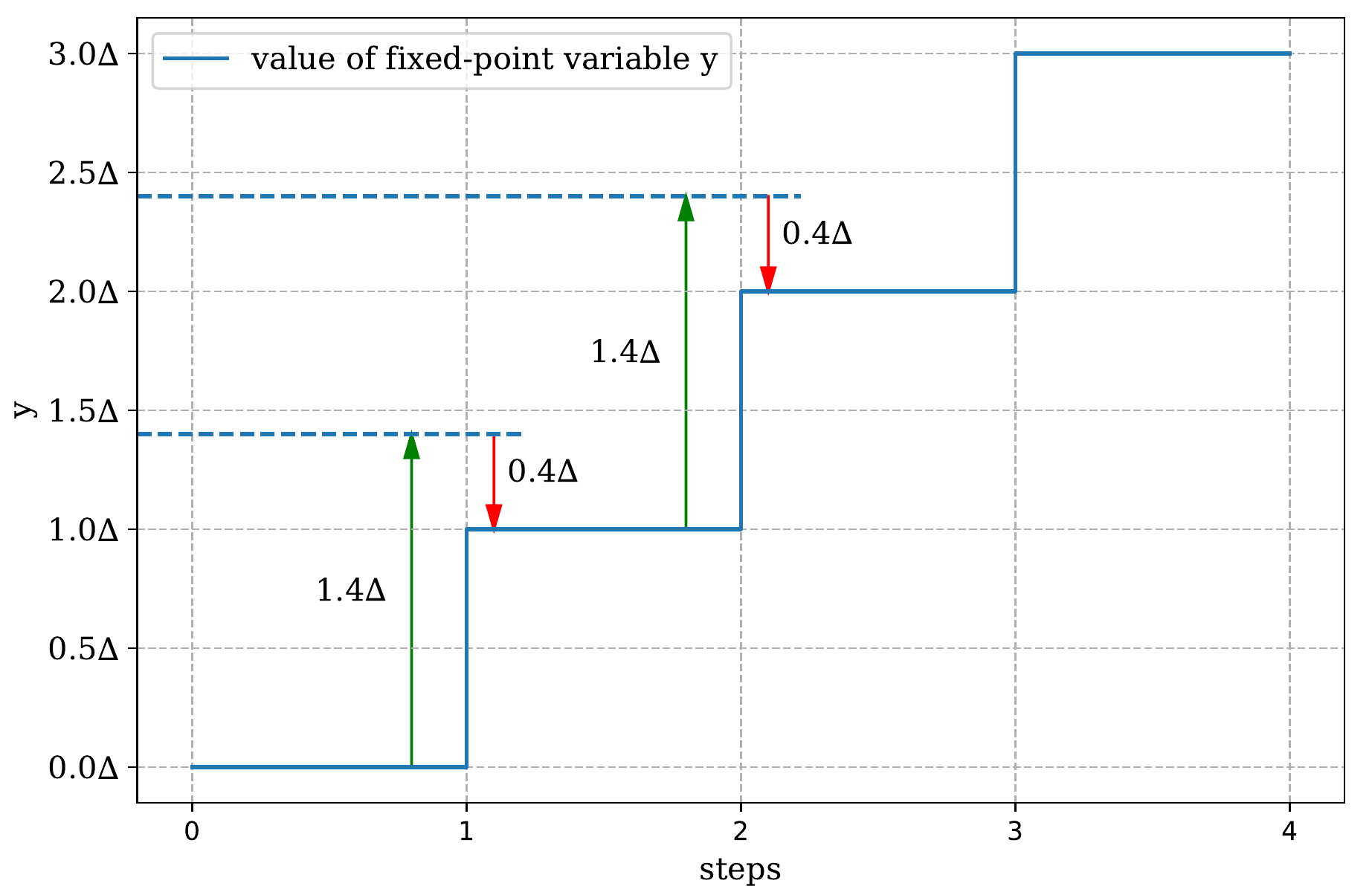}
  \caption{\changed{}{An illustration of the dependence between quantization errors. At each frame, the variable $y$ is added by $1.4\triangle$ (see the green arrows) and then $y$ is rounded down causing a quantization error of $-0.4\triangle$ (the red arrows). In this simulation pattern, the quantization errors for each frame are all equal to the previous one. Therefore, assuming they are independent random variables is not always reasonable.}}
  \label{fig:dithering_illustration}
\end{figure}

\changed{So far, we model the quantization error as random uniform distribution for each quantization operation and asserted uncorrelated between each operation.}{So far, we model the quantization error as random uniform distribution for each quantization operation, and we assume they are independent in the simulation.}
However, it is not a realistic assumption in general settings.
The quantization error is directly coupled with the digit after the least significant bit. It is liable to reflect the functional relationship between the state variables.\changed{, which may be caused by the iteration between steps or the interaction of adjacent particles.}{}
In other words, if the uncorrelated assumption on the quantization error $Cov(e_{t_1,i_1}, e_{t_2, i_1})=0 $ does not hold for every distinct pair of $(t_1, i_1) \neq (t_2, i_2)$, it is not safe to apply Eq.~\eqref{equ:error_prop}, since the actual error can be much bigger than the estimation.

\changed{}{As an example, assuming we have a variable $y$ represented by a two-bit unsigned fixed-point number with four quantization levels. Starting from $y=0$, we add $1.4\triangle$ to $y$ for each time step. 
In this pattern, the quantization error of the first step is $-0.4\triangle$, which leads to the subsequent quantization errors being all $-0.4\triangle$. Therefore we cannot assume the quantization errors are independent random variables in this case.
Please see Fig.~\ref{fig:dithering_illustration} for details.}



As shown in \cite{generalquant,quantspectra,amplitudequantized}, dithering is an effective solution to supply the necessary condition of modeling the quantization error as uniform and uncorrelated.
Overall, we follow the practice of non-subtractive dithering \cite{nonsubtractive} in our system.
To elaborate, in each store operation, we impose an additive uniform noise before rounding:

\begin{equation}
    u = \theta_{dither}(v) = \lfloor \frac{v}{\triangle} + \xi \rceil, \quad
    \xi \sim U(-\frac{1}{2}, \frac{1}{2})
\end{equation}



\begin{figure}
    \centering
    \begin{minipage}[t]{\linewidth}
      \centering
      \includegraphics[width=\linewidth, trim=246px 190px 93px 222px, clip]{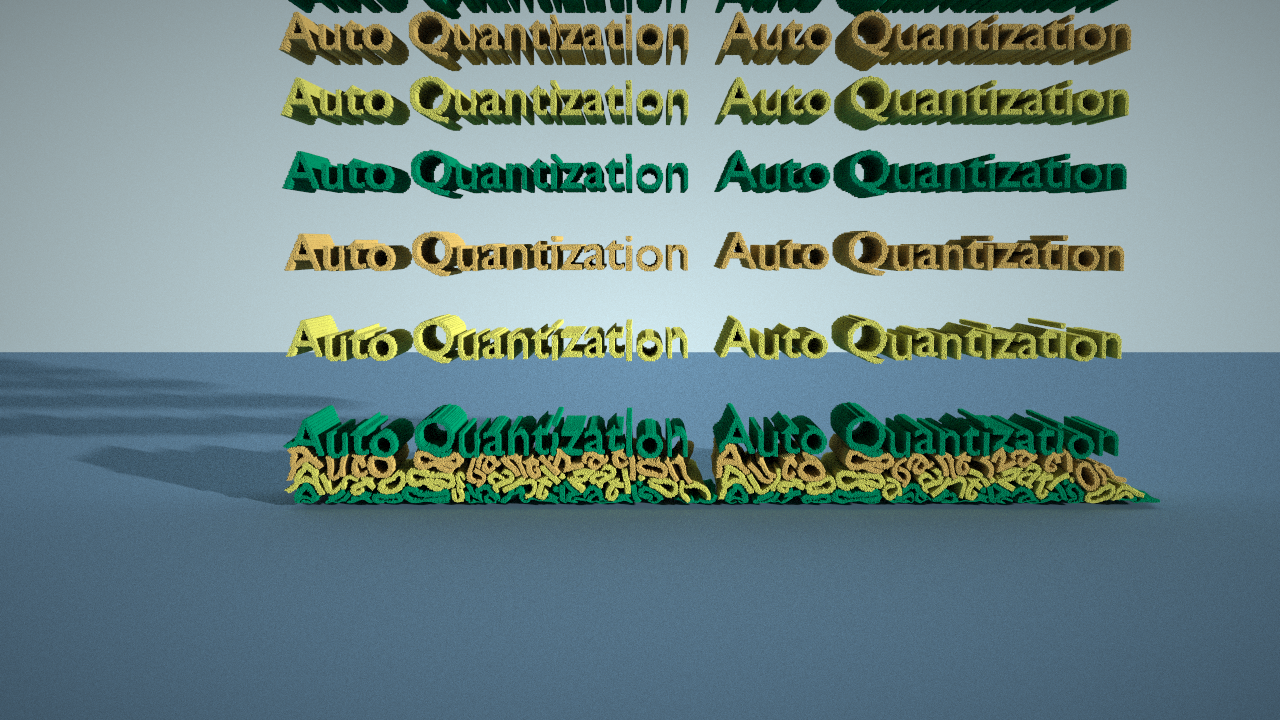}

      \vspace{-61pt}
      \hspace{-8pt}
      \includegraphics[width=0.9\linewidth]{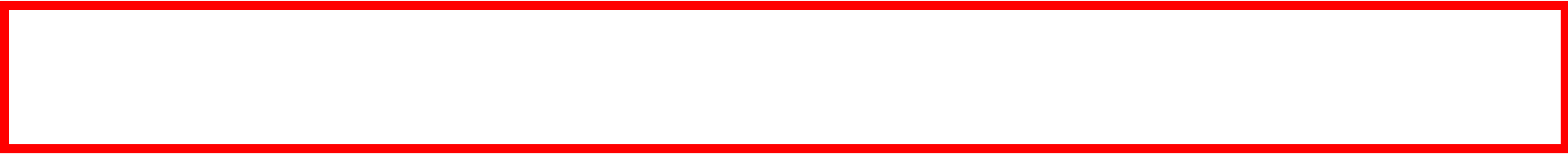}
    \end{minipage}

   \vspace{1pt}
    \begin{minipage}[t]{\linewidth}
      \centering
    \includegraphics[width=\linewidth, trim=246px 190px 93px 232px, clip]{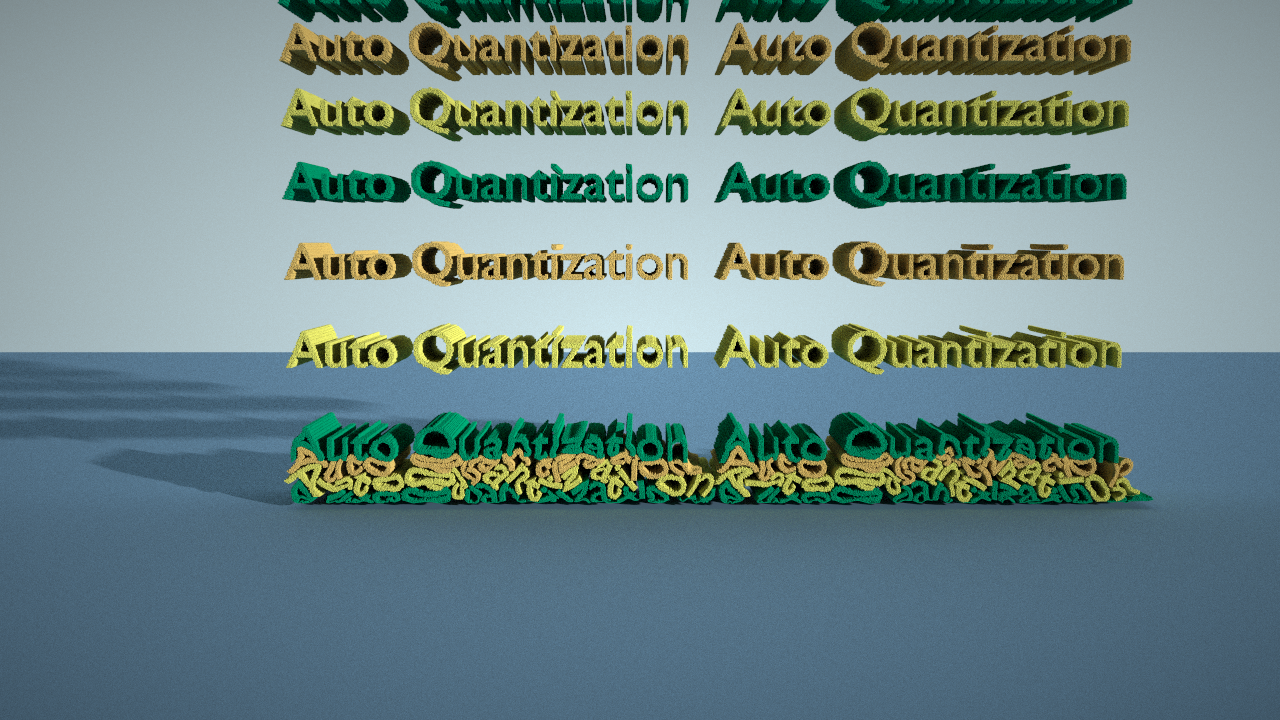}

      \vspace{-58pt}
      \hspace{-8pt}
      \includegraphics[width=0.9\linewidth]{figures/red_rectangle.pdf}
    \end{minipage}
  \caption{Effect of our dithering scheme. Top: quantized simulation \emph{without} dithering where the letters fall with different vertical velocities. Bottom: quantized simulation \emph{with} our dither scheme where the letters align neatly.}
\label{fig:dithering_compare}
\end{figure}

This dithered representation is a more accurate description than a deterministic round-to-nearest scheme because it has an \emph{unbiased} first moment.
Even with unevenly distributed signals, dithered quantization presents a stronger tolerance and leads to a less biased result.
Specifically, if we denote the normalized round-off error, i.e. $\frac{v}{\triangle} - \lfloor \frac{v}{\triangle} \rfloor$, as $Y \in [0,1)$, the probability of performing a round-up is exactly $Y$, i.e. $\mathscr{P}(Y + \xi>0.5) = Y$.
In other words, the dithered value after rounding shares the same expectation of $Y$ regardless of the distribution of $Y$.
In comparison, if $Y$ distributes unevenly between the two intervals $[0, 0.5),[0.5,1)$, a naive rounding scheme will result in a bias in the expectancy.
However, the error variance still depends on the value of $Y$ and the $\frac{1}{12}\triangle^2$ variance of quantization error remains an unsubstantiated assumption.
In practice, our dithering scheme works better than an alternative implementation without dithered quantization (see Fig.~\ref{fig:dithering_compare} for an example). 


\section{Gradient Computation}
\label{sec:gradient_computation}


\changed{}{Our method relies on the gradient of each variable involved in the estimation of the system error in Eq.~\eqref{eq:expectation_with_delta_h}}. \changed{We implement our algorithm based and won the differentiable programming framework proposed by DiffTaichi~\cite{hu2019difftaichi}}{We thus adopt the differentiable programming framework of DiffTaichi~\cite{hu2019difftaichi}}
and adhere to its restrictions to compute the gradients. \changed{Additionally, to support longer simulation duration, we follow the bisection algorithm. However, \cite{griewank1992} to achieve logarithmic spatial complexity with respect to growth in the number of steps.}{However, DiffTaichi requires storing the states of each frame for back-propagation. This requirement leads to linear growth of memory footprint, which restricts the steps of the simulation. Thus, we follow the bisection algorithm \cite{griewank1992} to achieve logarithmic spatial complexity w.r.t the growth of simulation steps.}

With auto-diff implementation, gradient $\frac{\partial z}{\partial \systemconfig_{t}}$ is calculated from the chain rule below:
\begin{align}
    \frac{\partial z}{\partial \systemconfig_t} = \frac{\partial z}{\partial \systemconfig_{t+1}} \frac{\partial \systemconfig_{t+1}}{\partial \systemconfig_{t}}
\end{align}
where $\frac{\partial z}{\partial \systemconfig_{t+1}}$ is the result from last step which is carried backward, and $\frac{\partial \systemconfig_{t+1}}{\partial \systemconfig_{t}}$ is a function of $\systemconfig_{t}$. In all, the compiler transforms the forward program $\systemconfig_{t+1} = \mathscr{F}(\systemconfig_{t})$ to the adjoint kernel $\mathscr{G}$ which calculates $\frac{\partial z}{\partial \systemconfig_{t}}$ from input $\systemconfig_{t}$ and $\frac{\partial z}{\partial \systemconfig_{t+1}}$:
\begin{align}
    \frac{\partial z}{\partial \systemconfig_{t}} = \mathscr{G}(\systemconfig_{t}, \frac{\partial z}{\partial \systemconfig_{t+1}})
\end{align}
Therefore, we must recover state $\systemconfig_{t}$ before execution of adjoint kernel before back-propagating the gradients from $\systemconfig_{t+1}$ to $\systemconfig_{t}$. In the implementation of ChainQueen, all the states $\systemconfig_{0}, \dots, \systemconfig_{T}$ are stored in pre-allocated areas so that they can be referenced out of the box \cite{hu2019chainqueen}.

Our bisection algorithm is similar to DiffTaichi's checkpointing method.
Each checkpoint stores the entire time section so that a rerun is possible.
Trading time for space is possible by not keeping all the states at hand but only recovering them by restarting from the nearest checkpoint when necessary.
However, instead of dividing the simulation into uniform segments as suggested by DiffTaichi, we allocate the checkpoints unevenly so that each checkpoint is in the middle of the last checkpoint and the current state.

\begin{figure}
    \centering
    \includegraphics[width=\linewidth]{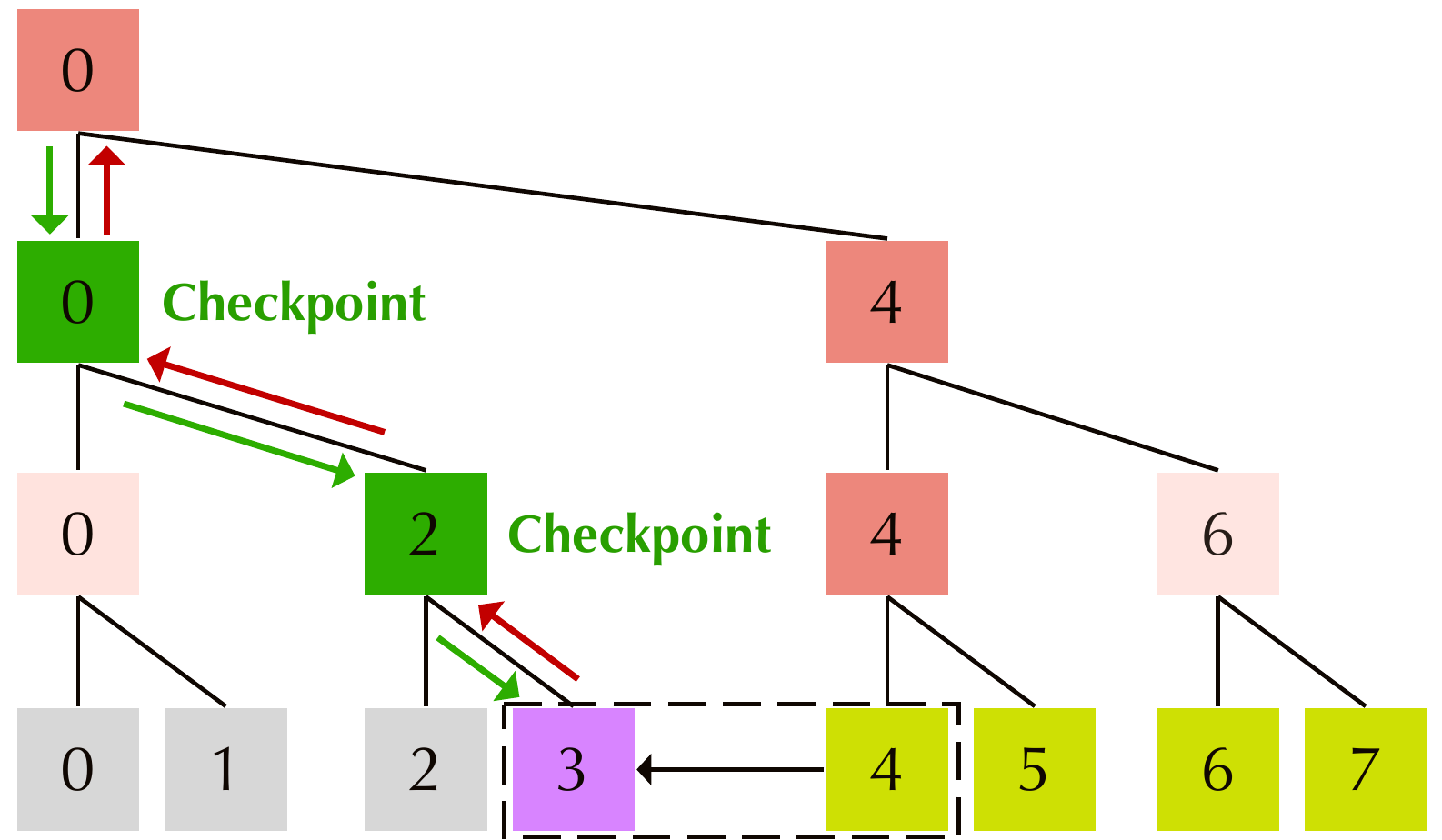}
  \caption{Illustration of the bisection algorithm. Adjoint kernel is back propagating $\frac{\partial z}{\partial \systemconfig_4}$ to $\frac{\partial z}{\partial \systemconfig_3}$. The states at steps 0,4 are stored as checkpoints (shown in red). To recover $\systemconfig_3$, retrace to the nearest non-empty checkpoint (red arrow), which is state $\systemconfig_0$, the nearest common ancestor of leaf node of state $\systemconfig_3$ and $\systemconfig_4$. A rerun is then executed from $\systemconfig_0$. During the rerun process, the checkpoints at depth 1 and 2 (with root at depth 0) are updated after 0 and 2 steps respectively (green arrow, new checkpoints shown in green).}
  \label{fig:bisection}
\end{figure}

In practice, we organize the states $\systemconfig_{0}, \dots, \systemconfig_{T}$ in a binary B+ tree, with storage only on its leaves.
The leaf nodes are indexed by $0,\dots, T$ from left to right, corresponding to the states $\systemconfig_{0}, \dots, \systemconfig_{T}$. Each non-leaf node serves as a signpost, which records the smallest value of their sub-trees.
We equate the checkpoints to the non-leaf nodes according to the value, i.e., a non-leaf node with a signpost value $t$ corresponds to a checkpoint storing the information of state $\systemconfig_{t}$.
In a time step $t$, only the checkpoints on the path from the root to leaf node $t$ reside in memory. 
Upon execution of the routine $\frac{\partial z}{\partial \systemconfig_{t}} = \mathscr{G}(\systemconfig_{t}, \frac{\partial z}{\partial \systemconfig_{t+1}})$ to back-propagate gradients from step $(t+1)$ to $t$, state $\systemconfig_{t}$ is required as an input.
With the above structure, we trace back to the nearest common ancestor of $\systemconfig_{t}$ and $\systemconfig_{t+1}$ where we will find the nearest checkpoint available and then execute a rerun from there.  (\changed{}{Please refer to  Fig.~\ref{fig:bisection} for an illustration.})

\begin{figure}
    \centering
    \includegraphics[width=\linewidth]{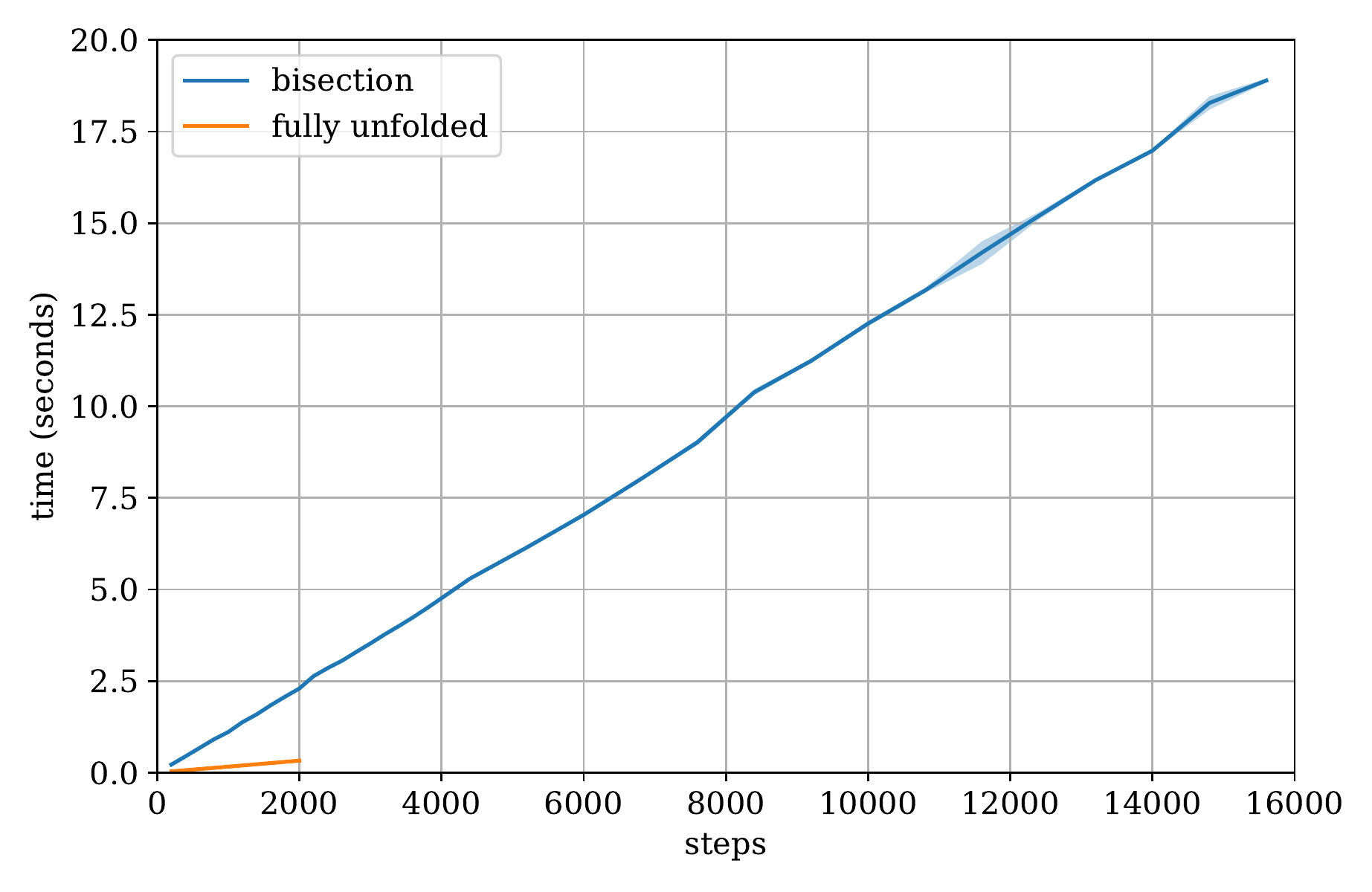}
    \caption{Comparison of temporal complexity between the bisection algorithm (blue) and the fully unfolded scheme (orange). Note that the orange line ends after 2000 steps since GPU fails to allocate enough memory, and the bisection algorithm is roughly 8 times slower under this situation. However, the bisection scheme is nearly free from memory limitations.}
    \label{fig:time_comlexity}
\end{figure}

Overall, this bisection algorithm achieves temporal and spatial complexity of $O(N \log N)$ and $O(\log N)$, respectively, as opposed to $O(N)$ and $O(\sqrt{N})$ of DiffTaichi, and thus significantly expanding the capacity of total number of steps in a memory-bound scenario.
The actual running time is reported in Fig.~\ref{fig:time_comlexity}.

\section{Bit Pack Data Structure}
\changed{To better enhance the productivity of our workflow,}{After solving the optimization problem defined in Eq.~\eqref{eq:sim_err} or Eq.~\eqref{eq:sim_mem}, we need to place the custom-length data types into the memory automatically. Therefore, }we propose a new bit-level data structure, namely \emph{bit pack}, which allows us to easily place arbitrary length custom data types into memory \changed{}{without worrying about the bit-level layout}.
Similar to the bit-level container \emph{bit struct} proposed by~\cite{hu2021quantaichi}, \emph{bit pack} is also a tree-based custom data type container.
However, \emph{bit struct} only supports hardware-native data types as the variable containers, which makes users restrained when working with custom data types.
There are two main limitations of \emph{bit struct}. First, users have to modify the code describing the placements of quantized data types if the number of bits inside the \emph{bit struct} exceeds the maximum length of the physical type.
The following code snippet is an example.
The code describes the memory usage of a particle in a 3D MPM simulation, where $\mathbf{p}$, $\mathbf{v}$, and $\mathbf{F}$ represent the position, velocity, and deformation gradient of the particle, respectively.

\begin{lstlisting}
# using bit struct
ti.root.bit_struct(num=64).place(p(0), p(1), p(2))
ti.root.bit_struct(num=64).place(v(0), v(1), v(2))
ti.root.bit_struct(num=64).place(F(0, 0), F(0, 1), F(0, 2), F(1, 0))
ti.root.bit_struct(num=64).place(F(1, 1), F(1, 2), F(2, 0), F(2, 1))
ti.root.bit_struct(num=32).place(F(2, 2))
\end{lstlisting}
\begin{figure}
    \centering
    \includegraphics[width=\linewidth]{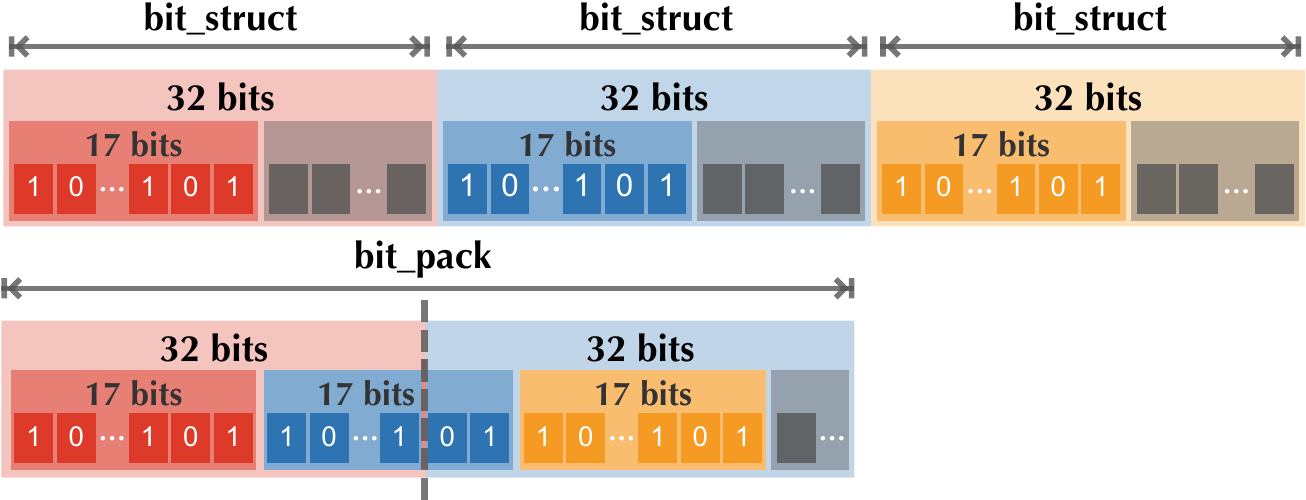}
  \caption{Comparison of \emph{bit pack} with \emph{bit struct}. Two 17-bit numbers do not fit into a single 32-bit \emph{bit struct}, and the three 17-bit elements consume three \emph{bit structs}. In contrast, the three elements can be contained by two 32-bit \emph{bit packs}.}
  \label{fig:bit_struct}
\end{figure}

\begin{figure}
    \centering
    \includegraphics[width=\linewidth]{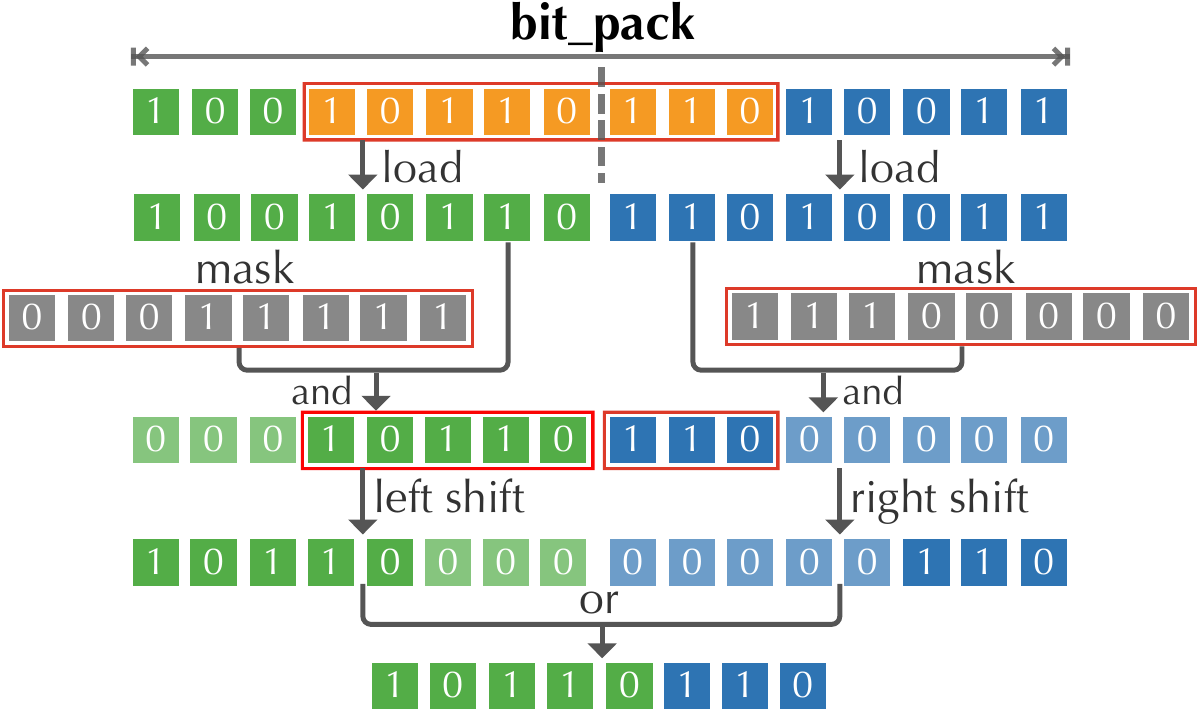}
  \caption{Illustration of the load operation on a custom type variable spanning two different physical words based on our \emph{bit pack} data structure. We first load the two words, then use a series of bit operations to reconstruct the original data.}
  \label{fig:bit_pack_operation}
\end{figure}

In the example above, the total bit width of $\mathbf{p}$ cannot exceed 64. However, this constraint cannot be guaranteed before the quantization scheme is derived from our optimization. Users may need to re-organize the placement, for instance, move the last component of $\mathbf{p}$ into another \emph{bit struct} if the bit width violates the limit.
Besides, when designing a quantized simulator manually, trial and error takes the most of the time. Frequently adjusting the placement by modifying codes can be troublesome.

The second limitation of \emph{bit struct} is that it does not allow custom data types to span across two physical words. See Fig.~\ref{fig:bit_struct} for an example. Supposing we have three 17-bit custom integers to be placed in 32-bit \emph{bit struct}s. Due to the limitation, we have to use $3\times32$ bits to accommodate them, leading to $3\times15$ bits unfilled.

We propose \emph{bit pack} to address this problem using a simple strategy.
Custom data types can be placed into the memory contiguously, and the compiler will detect the case that the data types span across two physical words.
When a data type is split by physical words, the compiler will generate code performing a series of bit operations to reconstruct the custom data types.
See Fig.~\ref{fig:bit_pack_operation} for a visual demonstration.
With the help of \emph{bit pack}, the above example only requires one line of code to describe the memory placement clearly as below, and no more code changes in other parts are needed.
\begin{lstlisting}
# using bit pack
ti.root.bit_pack().place(p, v, F)
\end{lstlisting}

\section{Experiments and Evaluations}
We conduct experiments to evaluate the effectiveness, performance, and scalability of our method.
We further develop three large-scale simulators with our workflow.
Implementation details of all the quantization schemes presented in this section can be found in the supplemental materials.

\subsection{Effectiveness}

We evaluate the controllability of our approach in both \emph{error-bounded} and \emph{memory-bounded} scenarios \changed{we develop three memory-intensive simulators including}{using three 2D simulators, including an MLS-MPM~\cite{hu2018moving} elastic body simulator, an MLS-MPM fluid simulator, and an advection-reflection~\cite{zehnder2018advection} Eulerian smoke simulator.}

\paragraph{Error-bounded quantization}
\label{sec:error_bound}
To evaluate the capability of the error-bounded quantization, we conduct experiments of 2D elastic body and Eulerian fluid simulation on an NVIDIA GTX 1080Ti GPU with 11 GB memory.

We simulate eight elastic squares falling down and colliding with each other based on the MLS-MPM algorithm~\cite{hu2018moving}.
Particles are the primary memory consumer, while background grids use less memory due to sparsity.
We store positions ($\mathbf{p}$), velocities ($\mathbf{v}$), deformation gradients ($\mathbf{F}$) and affine velocity field ($\mathbf{C}$) on each particle which add up to $48$ bytes with IEEE754 single floating-point precision.
The MPM experiment is conducted with $80,000$ particles and $128\times 128$ background grids.
We use the kinetic energy in the final state as the evaluation function.
The entire simulation process includes 8192 steps with a time interval of $2 \times 10^{-4}$.

Our 2D Eulerian Smoke experiment is conducted using a differentiable advection-reflection~\cite{zehnder2018advection} solver.
We perform advection in a semi-Lagrangian scheme with RK-3 path integration.
We solve Poisson's equation using $64$ Jacobian iterations.
The grid resolution and time interval are $256 \times 256$ and $0.01$, respectively.
The evaluation function is set to be the squared sum of smoke density on all grids in the final state.
Finally, the quantized variables are the pressure and velocity of the grids.

\begin{figure}
    \centering
    \includegraphics[width=\linewidth]{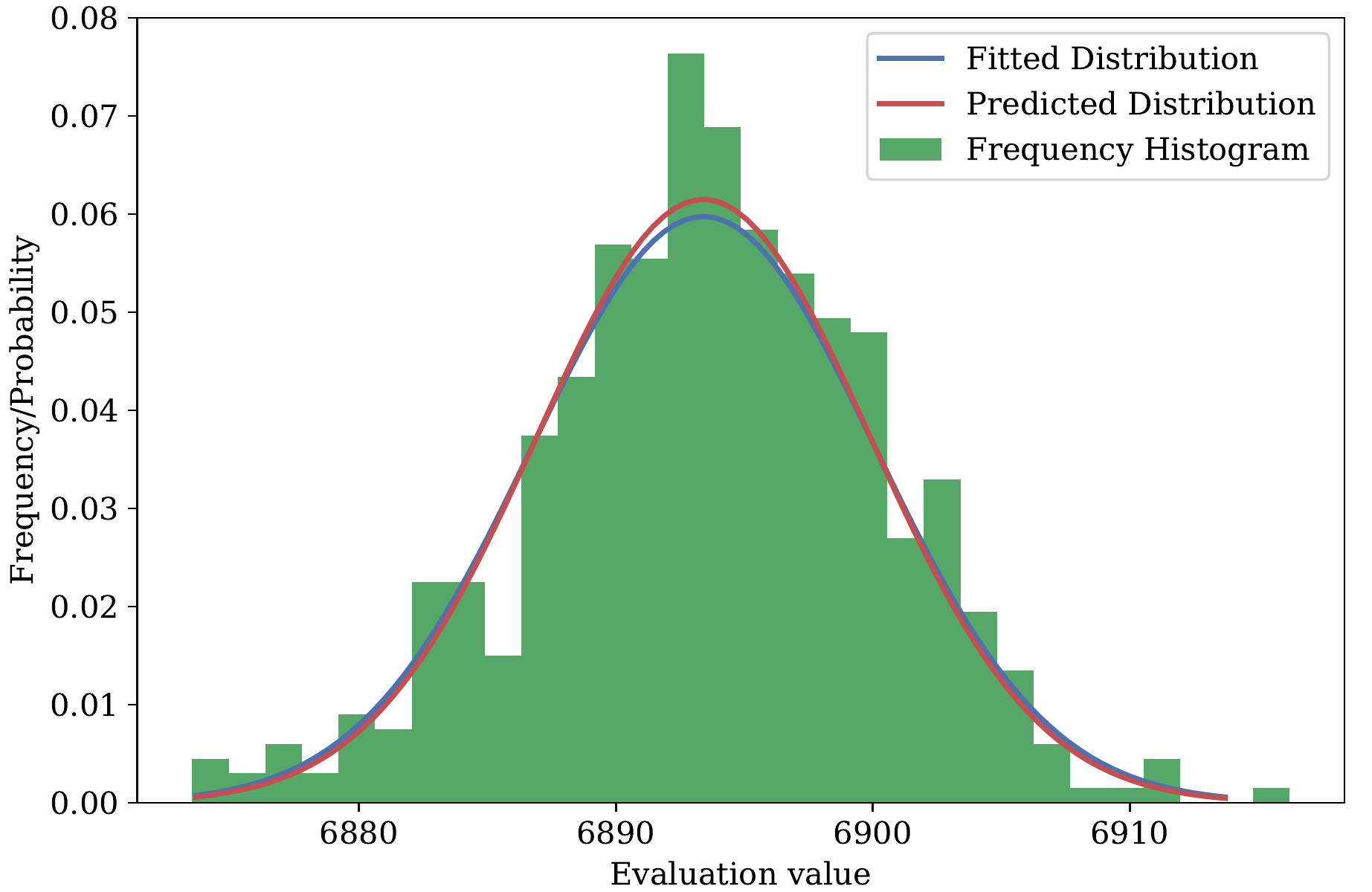}
  \caption{Distribution of the evaluation function on quantized MPM elastic body simulation. The probability density function fitted from the mean and variance of the observed samples is shown in blue. \changed{The red curve represents the predicted density function from the derivatives.}{The red curve representing the predicted density function is computed through Eq.~\eqref{eq:expectation_with_delta_h} with the gradients of the simulation and the testing quantization resolution.}
  The close match between the two curves validates our error estimation model.
  }
  \label{fig:distribution}
\end{figure}

In our experiments, we consider the evaluation function of a quantized simulation as a normal distribution.
To validate this assumption, we first run a quantized 2D MPM elastic body simulation 500 times and record the frequency of the function.
The histogram of the distribution is shown in Fig.~\ref{fig:distribution}, from which we can see the values of the evaluation function can be well approximated using a normal distribution. \changed{}{Please refer to the 4th column of Table 1 in our supplementary materials for the quantization scheme used in this experiment.}

We set the relative error tolerance $\epsilon_{err}$ to four different magnitudes: $10^{-1}$, $10^{-2}$, $10^{-3}$, and $10^{-4}$.
The quantized simulation is executed 20 times for each magnitude, where the mean and standard deviations are recorded as $\mu_{sim}, \sigma_{sim}$.
Since we model the evaluation function as a normal distribution, we consider the error is successfully under control if the evaluation function value of a quantized simulation falls in a range of $3\epsilon_{err}$ centered at $\mu_{sim}$. \changed{}{We record the number of successful controls out of 20 tests as the success rate.}

From Table~\ref{tab:error-comparsion} we can see that as the error tolerance increases, the number of quantization bits increases, and the fluctuation of the evaluation function becomes smaller.
The bias and variance satisfy well with the error tolerance in most cases.
The failure case of Eulerian smoke in the second row of Table~\ref{tab:error-comparsion} is quantized with only 4 fraction bits on pressure and 6 fraction bits on velocity. \changed{}{This shows the potential hazard of abuse, where a radical  setting of the error bound might lead to a bold quantization scheme and a large }deviation from the full-precision simulation.
In this situation, the local linearization cannot approximate Eq.~\eqref{equ:second_momentum} well enough.

\changed{}{
As mentioned above, our method can be solved iteratively, but we only adopt the solution without iteration for lower computational costs.
To validate this design choice, we solve the optimization iteratively in the example of 2D smoke simulation. We find that the optimization process converges after no more than six iterations, and the corresponding result of the iterative approach differs from the non-iterative solution by at most one bit on each variable.
However, the iterative solution is highly expensive due to the re-computation of gradients at each iteration, so we keep using the non-iterative solution in the rest of our experiments.
Please refer to section 3.1 in the supplementary material for the detailed results.
}

\begin{table}
\centering
\caption{Results of the error-bounded experiments. \changed{}{The MPM elastic body and Eulerian smoke simulations are labeled as Exp. 1 and Exp. 2. The column of \emph{$\epsilon_{err}$} is the relative error tolerance used in each experiment. \emph{Control Objective} in the table }dictates the absolute tolerance of the simulation error. \changed{}{We use \textbf{S.R.} and \textbf{M.C.} as abbreviations for success rate and memory compression.} The quantized evaluation function lands nicely within the range of permitted error bound. Of all the $160$ experiments, there are $136$ times that our quantization scheme manages to control the value.}
       \resizebox{\linewidth}{!}{%
    \begin{tabular}{c|c|c|c|c|c}
    \toprule
    \textbf{Exp.}     &$\bm{\epsilon_{err}}$ & \textbf{ Control Objective } & \textbf{\begin{tabular}[c]{@{}c@{}}Quantized\\ Evaluation Function\end{tabular}} & \textbf{S.R.} & \textbf{M.C.} \\ \midrule \midrule
    \textbf{1}    &0.1& $6.9 \pm 2.1 \times 10^3 $                        & $7.1  \pm 1.1 \times 10^3 $       & 19/20     & $2.00\times$  \\
    \textbf{2}   &0.1& $1.87\pm 0.56\times 10^4 $                        & $3.69 \pm 0.08 \times 10^4 $         & 0/20   & $5.05\times$ \\ \hline
    \textbf{1}    &0.01& $6.89 \pm 0.21\times 10^3 $                      & $6.90 \pm 0.06 \times 10^3 $         & 20/20  & $1.67\times$ \\
    \textbf{2}   &0.01& $1.875\pm 0.056\times 10^4 $                     & $1.90 \pm 0.04 \times 10^4 $         & 17/20  & $3.20\times$ \\ \hline
    \textbf{1}     &0.001& $6.893 \pm 0.021\times 10^3 $                   & $6.895 \pm 0.008 \times 10^3 $        & 20/20   & $1.42\times$ \\
    \textbf{2}   &0.001& $1.8749\pm 0.0056\times 10^4 $                  & $1.875 \pm 0.003 \times 10^4 $         & 20/20   & $2.40\times$ \\ \hline
    \textbf{1}     &0.0001 &$6.8934 \pm 0.0021\times 10^3$         & $6.8946 \pm 0.0007 \times 10^3 $       & 20/20   & $1.23\times$ \\
    \textbf{2}   &0.0001 &$1.87486\pm 0.00056\times 10^4$        & $1.87488 \pm 0.00019 \times 10^4 $     & 20/20   & $1.96\times$ \\ \bottomrule
    \end{tabular}
    }
   \label{tab:error-comparsion}
\end{table}

\paragraph{Memory-bounded quantization.}
We evaluate the scenario of memory-bound quantization on a 2D MLS-MPM based fluid simulation of a dam break example.
Similar to the previous MPM elastic body simulations in \emph{error-bounded} experiments, we store $\mathbf{p}$, $\mathbf{v}$, and $\mathbf{C}$ of each particles.
As for the deformation gradient $\mathbf{F}$, we follow Tampubolon et al.~\shortcite{tampubolon2017multi} to track its determinant $J$ instead of its matrix elements for improved numerical stability.
In a full-precision simulation, it takes $36$ bytes to store the attributes of each particle.
We use the kinetic energy of the final time step as the evaluation function.
We run the simulation with $10,000$ particles and $128\times 128$ background grids.
The simulation lasts for 16,384 steps with a time interval of $2\times 10^{-4}s$.
The experiments run on an NVIDIA RTX 3090 GPU with 24 GB memory.

\begin{figure}
\includegraphics[width=\linewidth]{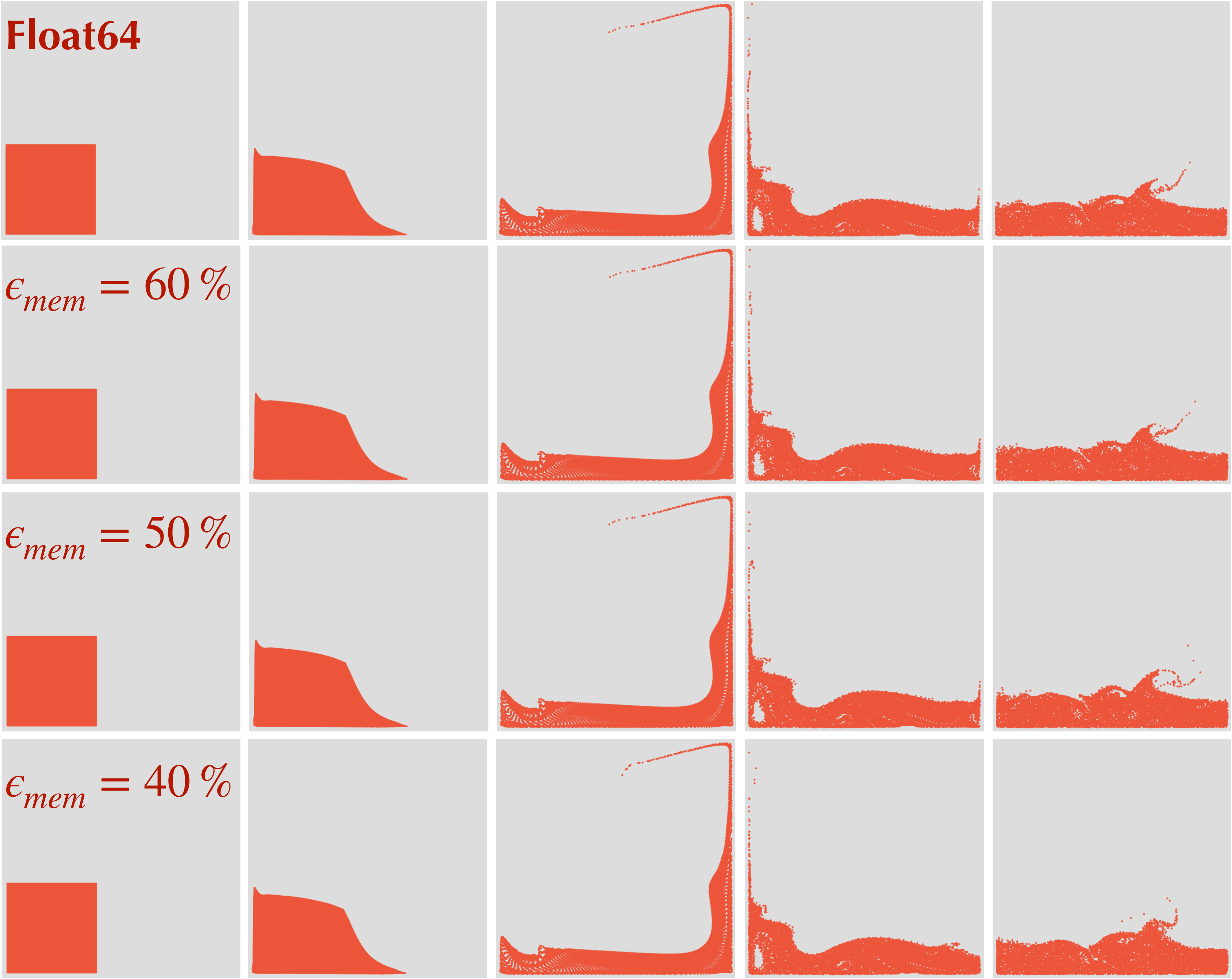}
  \caption{Visual comparison for the memory-bounded experiments.
  }
  \label{fig:compare_cr}
\end{figure}

We set the control parameters of the target compression rate from 40\% to \changed{70\% ($1.43\times$}{60\%}
We run the simulation 10 times for each target compression rate and record the mean $\mu_{mem}$ and standard deviation $\sigma_{mem}$.
The results can be found in Table~\ref{tab:compression-rate}.
Using the compression rate as the quantization target makes it difficult to define a successful error control as in the previous section, because the error bound is used as a soft constraint. So we use an error metric denoted as $e_{mem} = \sqrt{\sigma_{mem}^2 + (\mu_{mem} - z)^2}$, where $z$ is the reference value obtained by full-precision simulation.
Given this error metric, even in the case of the highest compression rate ($2.5\times$), the relative error is only $9.9$\%.
Meanwhile, the visual quality is still comparable to the full-precision version. Please refer to the visual comparison in Fig.~\ref{fig:compare_cr}.
In summary, our automatic quantization scheme successfully assigns appropriate numbers of quantization bits to each variable according to their contribution to the overall error under the limited memory budget.
\begin{table}
\centering
\caption{Memory-bound experiments. The target compression rate is the control parameter specified by users. The real compression rate is computed according to the quantization scheme derived from our optimization. The reference evaluation function value is $1.32 \times 10^3$.}
\resizebox{\linewidth}{!}{%
\begin{tabular}{c|c|c}
\toprule
 \textbf{\begin{tabular}[c]{@{}c@{}}Target \\ Compression Rate ($\epsilon_{mem}$)\end{tabular}} & \textbf{\begin{tabular}[c]{@{}c@{}}Real \\ Compression Rate\end{tabular}} & \textbf{Evaluation Function}  \\ \midrule \midrule
 60\% ($1.67\times$) & 58.3\% ($1.71\times$) & $1.28 \pm 0.06 \times 10^3$ \\
 50\% ($2.0\times$)  & 47.2\% ($2.12\times$) & $1.29  \pm 0.08 \times 10^3$ \\ 
  40\% ($2.5\times$) & 37.8\% ($2.64\times$)  & $1.25 \pm 0.11 \times 10^3$  \\ \bottomrule
\end{tabular}
}%
\label{tab:compression-rate}
\end{table}


\begin{table}
\centering
\caption{The success rate and evaluation function values for the first two experiments of optimality check. The base quantization scheme with a success rate of (20/20) and evaluation function of ($6.90 \pm 0.06 \times 10^3 $)  can be found in the supplemental material. }
\resizebox{\linewidth}{!}{%
\begin{tabular}{c|cc|cc}
\toprule
\multirow{2}{*}{\textbf{Bits Taken}} & \multicolumn{2}{c|}{\textbf{Success Rate}}                                          & \multicolumn{2}{c}{\textbf{Evaluation Function}} \\ \cline{2-5} 
                               & \multicolumn{1}{c|}{\textbf{Reduce All}} & \multicolumn{1}{c|}{\textbf{Reduce Half}} & \multicolumn{1}{c|}{\textbf{Reduce All} } & \textbf{Reduce Half} \\ \midrule
\textbf{1}                     & \multicolumn{1}{c|}{19/20}                 & 20/20                                     & \multicolumn{1}{c|}{$6.9 \pm 0.1 \times 10^3$}           &   $6.88 \pm 0.09 \times 10^3$         \\
\textbf{2}                     & \multicolumn{1}{c|}{13/20}                 & 17/20                                      & \multicolumn{1}{c|}{$6.9 \pm 0.3 \times 10^3$}           &     $6.8 \pm0.2 \times 10^3$       \\
\textbf{3}                     & \multicolumn{1}{c|}{5/20}                 & 12/20                                      & \multicolumn{1}{c|}{$7.3\pm 0.7 \times 10^3$}           &    $6.9\pm 0.4 \times 10^3$        \\ \bottomrule
\end{tabular}
}%
\label{tab:vicinity-search}
\end{table}

\begin{figure}
  \centering
  \includegraphics[width=\linewidth]{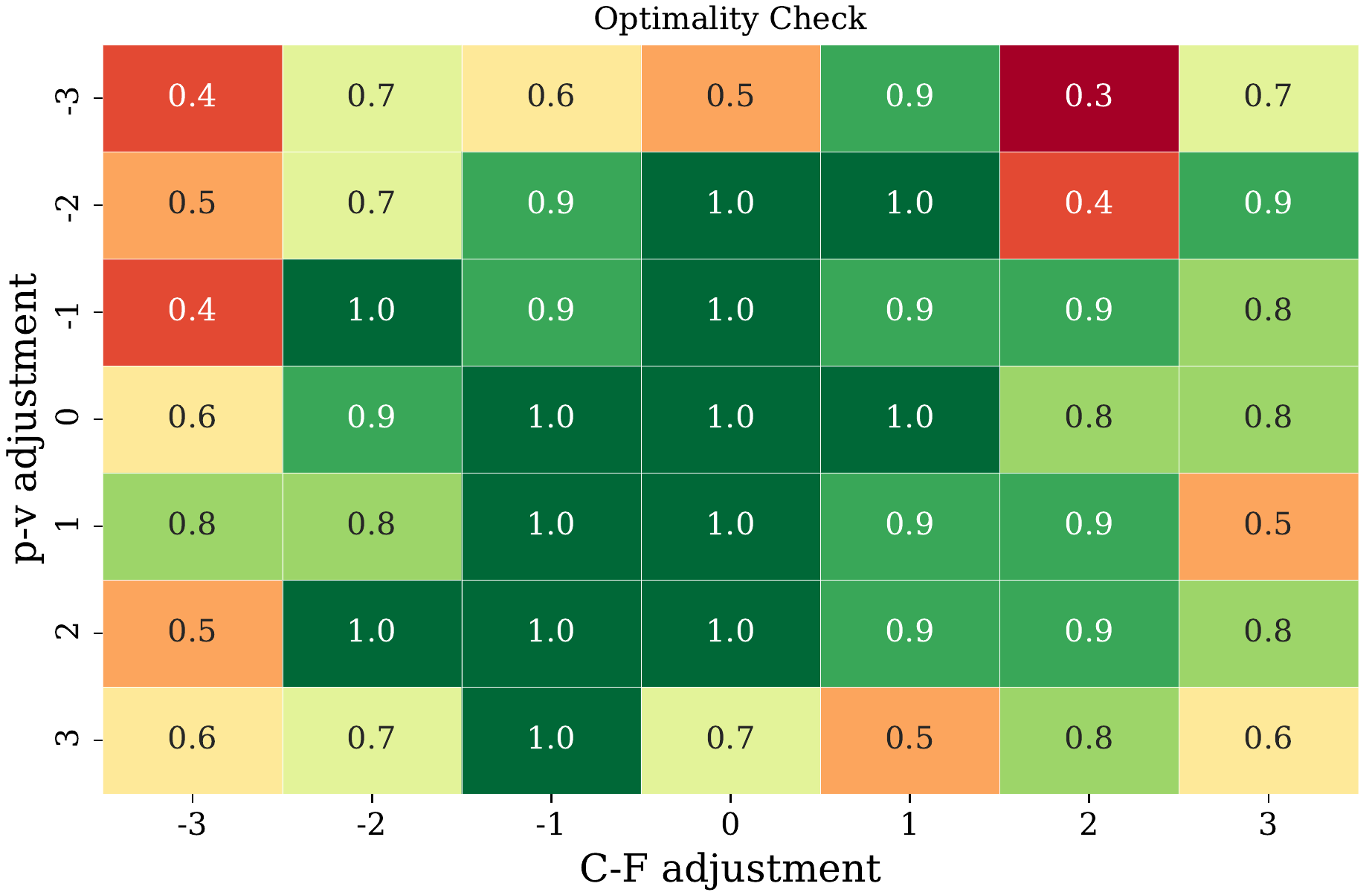}
  \caption{Optimality check. We conduct two orthogonal adjustments: to move bits from $\mathbf{p}$ to $\mathbf{v}$ (positive y-axis and vice versa) and from $\mathbf{C}$ to $\mathbf{F}$ (positive x-axis). The success rate is counted based on ten repetitions on each quantization scheme.
  }
  \label{fig:vicinity_search_fig}
\end{figure}


\paragraph{Optimality check}
Aided by the accurate prediction of the simulation error, the quantization scheme derived from our system can reach a very close vicinity to the critical point. We verify this claim using the previous MLS-MPM elastic body simulation with the same experiment settings. We first solve the optimization by setting the relative error bound to 0.01 and then make minor adjustments to the quantization scheme in three different ways:
\begin{enumerate}
    \item Unanimously take away bits in all data types.
    \item Randomly reduce the length on half of the data types. 
    \item Move bits from some data types to others.
\end{enumerate}


The bit reduction results (the first two methods) are recorded in Table~\ref{tab:vicinity-search}, and the result of bits-redistribution (the last method) is shown in Fig.~\ref{fig:vicinity_search_fig} and our supplemental materials. In these experiments, the success rates are counted by 20 and 10 repetitions, respectively.
This experiment reveals that even with less than 10\% modification of our generated scheme, there is a significant impact on the success rate of the error constraints, and our quantization scheme is almost the most memory-efficient scheme that satisfies the error constraints.

\begin{figure}
\includegraphics[width=\linewidth]{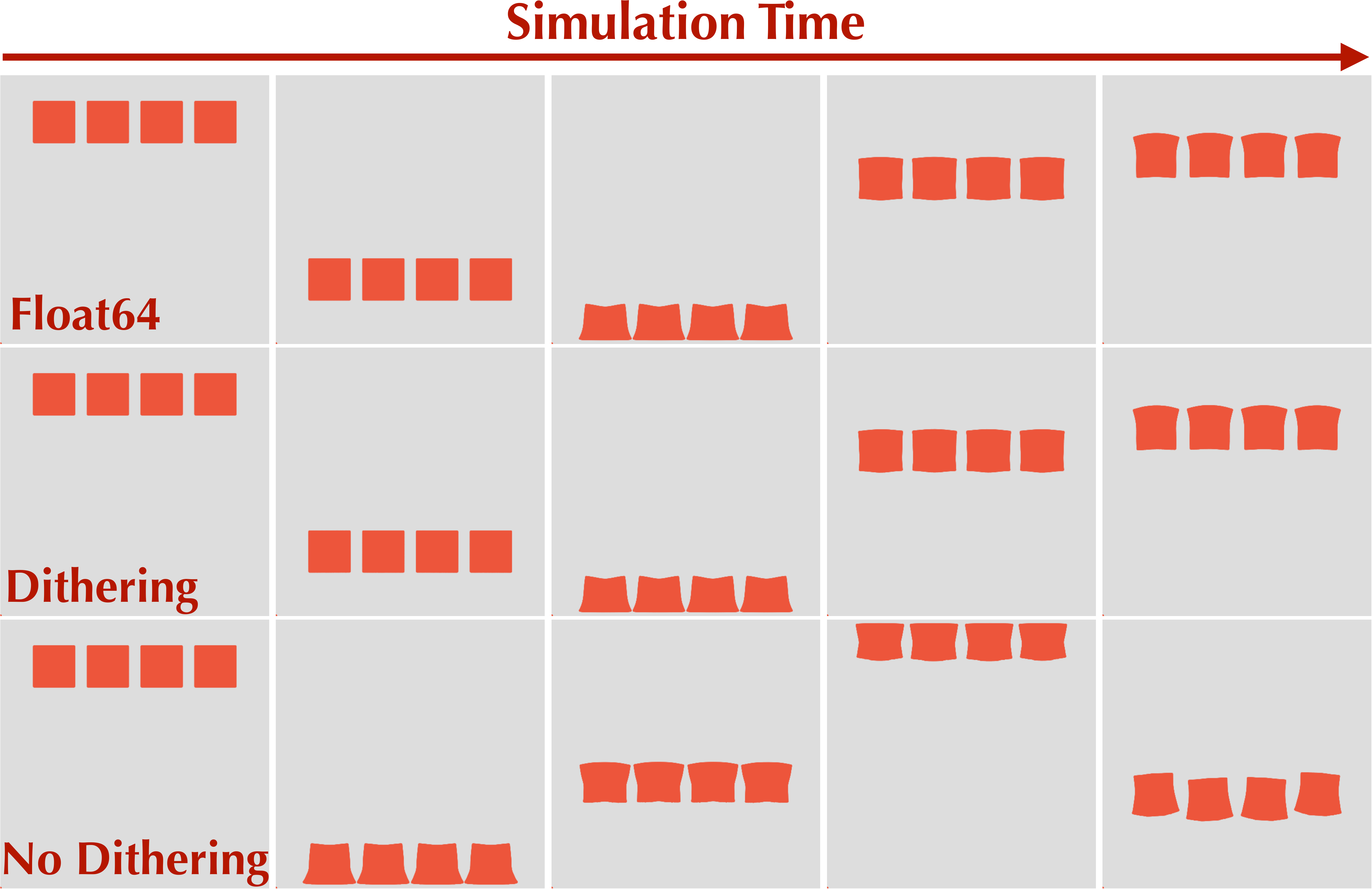}
  \caption{
  Effect of our dithering scheme.
  Without dithering (the bottom row), the elastic cubes fall from rest, bounce over the initial position, and smash the ceiling.
  In comparison, the dithered simulation with the same bit number (the second row) closely resembles the full-precision reference (the first row).}
  \label{fig:compare_dithering}
\end{figure}

\paragraph{Dithering}
To showcase the effectiveness of dithering in reducing the overall errors of quantized simulation, we use a quantization scheme with a relatively high memory compression ($2.5\times$). \changed{In line with previous sections, w}{}We choose kinetic energy as the evaluation function to measure the errors. Table~\ref{tab:dithering-effective} shows that the dithered simulation leads to a much more accurate evaluation function value in this challenging case, while the simulator exhibits significant deviation from the full-precision \floatpoint{64} simulation without dithering, which is confirmed by the target function value and the visual comparisons in Fig.~\ref{fig:compare_dithering}.
The explanation for the failure is that the induced round-off errors are ill-distributed, contradicting our assumption of uniform independence distribution in the error model.
We record the aggregated ratio of round-ups and round-downs and find that the ratio of rounding up and rounding down is $1.47$ for the $y$ component of velocity, as opposed to the dithered result of $1.00$ (See Table~\ref{tab:dithering-effective}). 

\begin{table}
\centering
\caption{The effectiveness of dithering. We compare three data types on a 2D MPM elastic body simulation. The result without dithering deviates far away from the full-precision reference and the case with dithering. Also the uneven ratio of round-ups to round-downs can be observed in the case without dithering.}
\resizebox{\linewidth}{!}{%
\begin{tabular}{c|c|c}
\toprule
\textbf{Data Type}         & \textbf{Evaluation Function} & \multicolumn{1}{c}{\textbf{\begin{tabular}[c]{@{}c@{}}Round-ups/\\ Round-downs\end{tabular}}} \\ \midrule \midrule
\floatpoint{64}                   &  $1.378 \times 10^5$ & -  \\ \hline
Quantize with dithering    &  $1.370 \times 10^5$ & 0.999976  \\ \hline
Quantize without dithering &  $8.885 \times 10^5$ & 0.683994   \\ \bottomrule
\end{tabular}
}
\label{tab:dithering-effective}
\end{table}

\begin{figure}
\centering
\begin{minipage}{\linewidth}
\includegraphics[width=\linewidth]{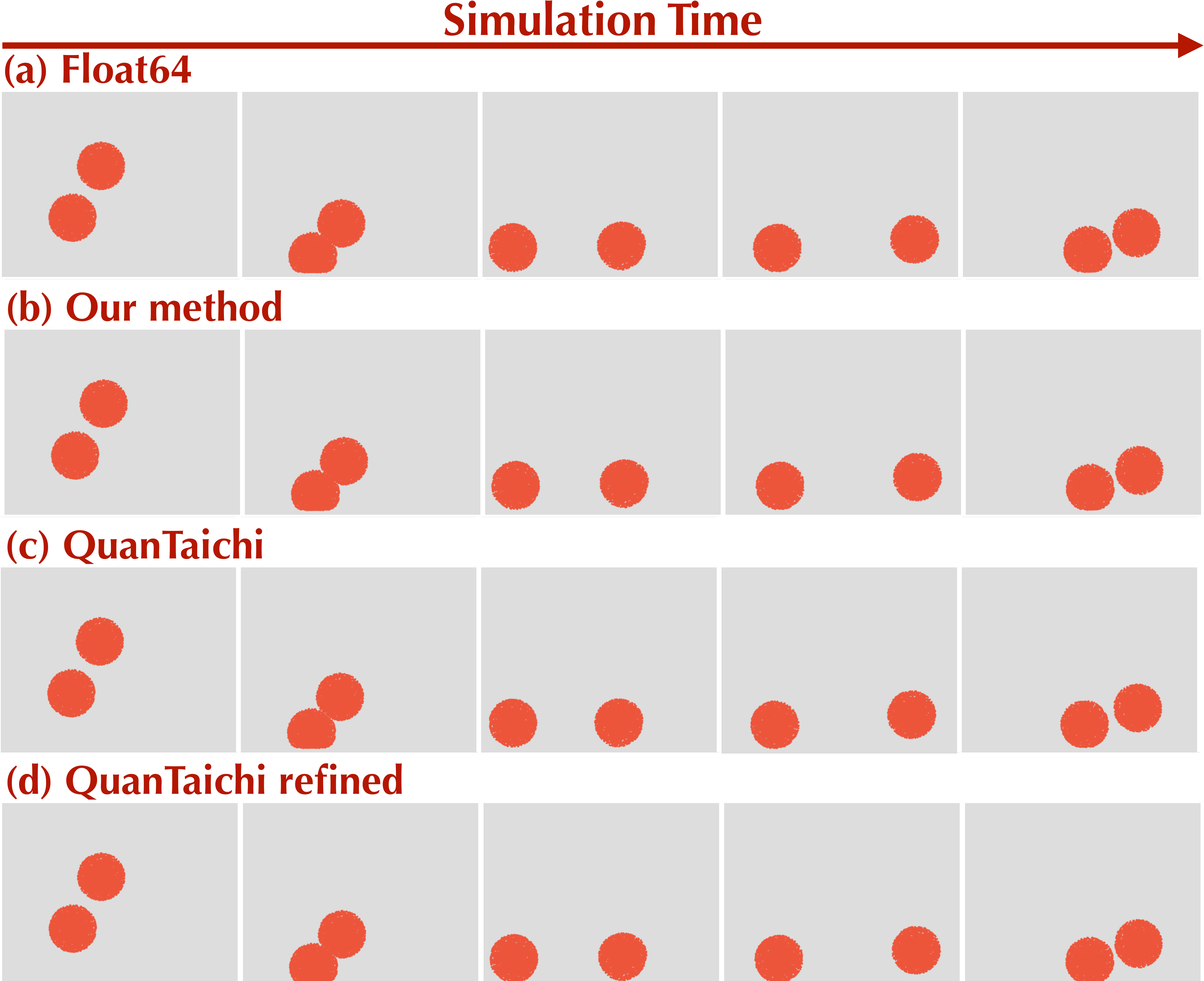}
\hspace{-33pt}
\includegraphics[height=0.7\linewidth]{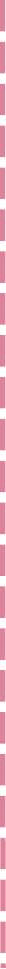}
\end{minipage}
\caption{\changed{}{The comparison with human generated schemes.
From top to bottom: (a) the \floatpoint{64} reference, (b) the result of our method, (c) the result of QuanTaichi~\cite{hu2021quantaichi}, and (d) the result via a refined version of (c).As we can see from the last frame, our scheme and the refined human-generated scheme are closer to the \floatpoint{64} reference, while our scheme uses $12.7$\% less faction bits.
The vertical dashed line is added to help compare the horizontal positions of circles by different methods.}
} 
\label{fig:compare_quantaichi}
\end{figure}
\changed{}{
\paragraph{Comparison with hand-tuned schemes}
We compare our method with human-generated results by reproducing the MPM ball collision experiment in QuanTaichi~\cite{hu2021quantaichi} and present the visual results in Fig.~\ref{fig:compare_quantaichi}.
We keep using the kinetic energy as the evaluation function.
All the components of position, velocity and deformation gradients are quantized in the experiment.
We first evaluate the error of the hand-tuned scheme $10$ times and use the mean error as the constraint. As a result, the quantization scheme derived from our optimization has $12.7$\% less number of fraction bits and $94.5$\% more minor error than the human-generated schemes. In the human-generated scheme, the velocity is represented by a custom float type with $6$ exponent bits and $10$ fraction bits. This data type has a larger dynamic range but has lower precision in a specific small range compared to the fixed-point data type. After we change it to a fixed-point data type and apply the ranges of our scheme for all variables, the error becomes $97.2\%$ smaller, which is very close to our scheme (see the last row of Fig.~\ref{fig:compare_quantaichi}).
The results show that our method can generate quantization schemes comparable to or even better than human experts.
}

\begin{table}
\centering
\caption{The benchmark of dithering performance on both CPU and CUDA backend. The quantization scheme can be found in the supplementary materials.}
\begin{tabular}{c|c|c|c}
\toprule
\textbf{Case} & \textbf{Backend} & \textbf{Dithering} & \textbf{No Dithering} \\ \midrule \midrule
\multirow{2}{*}{\textbf{Store}}          & x64                               & 54.516s                            & 8.888s                                \\ \cline{2-4} 
                                         & CUDA                              & 0.144s                             & 0.059s                                \\ \hline
\multirow{2}{*}{\textbf{MatMul}}         & x64                               & 115.888s                           & 16.593s                               \\ \cline{2-4} 
                                         & CUDA                              & 
                                         0.655s                             & 
                    0.262s                                \\ \hline
\multirow{2}{*}{\textbf{MPM 2D}}         & x64                               &  89.892s                                 &   59.498s                                     \\ \cline{2-4} 
                                         & CUDA                              & 26.819s                           & 26.651s                              \\ \hline
                                         
\changed{}{\textbf{MPM 3D}}         & \changed{}{CUDA} & \changed{}{ 29.094s }                                &  \changed{}{ 27.284s}                                     \\   
                                         \bottomrule
\end{tabular}
\label{tab:dithering_performance}
\end{table}

\subsection{Performance}
The techniques we presented in previous sections, such as dithering and \emph{bit pack}, may have impact on the performance.
We conduct several experiments to evaluate the performance impact of these techniques.
Unless with further explanation, all the experiments in this section are evaluated on an NVIDIA RTX 3090 GPU with 24 GB memory and Intel Core i7-7700k with 24 GB memory.
\paragraph{Performance benchmark}
We make a benchmark suite consisting of the following four different tasks to evaluate the running time:
\begin{description}
    \item[Store.] We record the time consumption of storing 512M 16-bit fixed-point numbers in to the memory ten times.
    \item[MatMul.] Similar to the former settings, we allocate 256M $3\times3$ matrices and multiply each with another $3\times3$ matrix ten times. Each element of the matrix is a 16-bit fixed-point number.
    \item[MPM 2D.] We test performance on a 2D MPM simulator with 8,000 particles, $128^2$ grids and 8,192 steps for five times.
    \item[MPM 3D.] \changed{}{We further test the performance on a more practical application of 3D MPM simulation with 30M particles, $1024^3$ sparse grid and 12,800 steps.}
\end{description}
We test the first three cases on both CPU and GPU backends and record the total time cost.\changed{}{
However, for the 3D MPM example, we only test it on GPU backend because CPU is not suitable for this large-scale simulation in practice. We record the running time per frame for this case. The quantization schemes of the 2D/3D MPM test cases can be found in section 2 of supplementary material.}
\paragraph{Dithering}
Dithering in Sec.~\ref{sec:dithering} introduces an extra random number generation step \changed{,}{that degrades the performance.}
\changed{We make a micro-benchmark suite consisting of three different tasks to evaluate the running time. 1) \emph{Store}: We allocate 256M 32-width bit-level containers and place two 16-bit custom float data into each of them. Then store the values into the memory ten times and record the time consumption. 2) \emph{MatMul}. Similar to the former settings, we allocate 256M $3\times3$ matrices and multiply each with another $3\times3$ matrix ten times. Each element of the matrix is a 16-bit fixed-point number. 3) \emph{MPM 2D}. In our implementation, the dithering procedure only takes place in the encoding stage of quantization operation, which, in an actual simulation program, only takes up a minor fraction in all the operations. We run a 2D MPM simulator five times with warm-ups to evaluate the performance loss in a real-world computation setting.}{
We test the influence of dithering using the benchmark
and the}
corresponding results can be found in Table~\ref{tab:dithering_performance}.
Our experiments reveal that storing custom \lstinline{float} types with dithering is a time-consuming operation, especially on a CPU backend, which slows down a memory-dense task by a factor of $8$ in the worst case.
\changed{}{Fortunately, the performance dose not significantly decrease on CUDA and the overall simulations of the MPM 2D/3D test cases are only $1.04\times$ slower on average as shown in Table.~\ref{tab:dithering_performance}.}
\changed{}{We further implement a faster random number generator to accelerate the dithering, improving the speed by $3.83\times$ on average (see Section 3.4 in the supplementary material for details).}

\begin{table}
\centering
\caption{The benchmark of \emph{bit pack} performance on CPU and CUDA backend. The quantization schemes and the data placement of the \emph{bit struct} are shown in the supplementary materials.}
\begin{tabular}{c|c|c|c}
\toprule
\textbf{Cases} & \textbf{Backend} & \textbf{Bit pack} & \textbf{Bit struct} \\ \midrule \midrule
\multirow{2}{*}{\textbf{Store}}          & x64                               & 7.173s                            & 7.376s                                \\ \cline{2-4} 
                                         & CUDA                              & 0.105s                             & 0.084s                                \\ \hline
\multirow{2}{*}{\textbf{MatMul}}         & x64                               & 13.608s                           & 18.269s                               \\ \cline{2-4} 
                                         & CUDA                              & 0.159s                             & 0.316 s                                \\ \hline
\multirow{2}{*}{\textbf{MPM 2D}}         & x64                               &  87.614s                                 &   112.109s                                     \\ \cline{2-4} 
                                         & CUDA                              & 26.155s                           & 26.814s                              \\ \bottomrule
 \changed{}{\textbf{MPM 3D}}         & \changed{}{CUDA} & \changed{}{ 29.094s }                                &  \changed{}{ 33.232s}                                     \\ 
 \bottomrule
\end{tabular}
\label{tab:bit_pack_perform}
\end{table}

\paragraph{Bit pack}
Using \emph{bit pack} will lead to cases where custom data types span across two adjacent physical words, introducing one extra memory access and several bit-level operations of shifting and masking.
To evaluate the performance impact, we use the performance benchmark described above with slight modifications on the quantization scheme.
Specifically, we use 20-bit and 25-bit fixed-points, in the \emph{Store} and \emph{MatMul} cases, respectively, to create situations of custom data types across two physical words.
A similar memory storage optimization strategy in QuanTaichi~\cite{hu2021quantaichi} is used to avoid atomicRMW when using \emph{bit pack}.
We compare the performance with \emph{bit struct} in Table~\ref{tab:bit_pack_perform}. \changed{}{To comprehensively evaluate the performance difference between the two data structures, we turn off dithering for the first two test cases and enable it for the two MPM cases.}
As can be seen from this table, the performance of \emph{bit pack} is comparable to or even better than \emph{bit struct} in most cases, since \emph{bit pack} leads to less memory consumption on both space and bandwidth with limited disruption to spatial locality.


\paragraph{Bisection algorithm}

We evaluate the performance of our bisection algorithm introduced in Sec.~\ref{sec:gradient_computation} on a 2D MPM solver with $5000$ particles and $128^2$ grids.
We record the average running time of a forward simulation and a back-propagation step combined, with eight repeations and three warm-up iterations, on an NVIDIA GTX 1080Ti GPU with 11GB memory.
The GPU can only contain 2000 states with a fully unfolded scheme before running out of memory, while the bisection algorithm enjoys \changed{a much}{at least $7.8\times$} larger carrying capacity in the experiment. See Fig.~\ref{fig:bisection} for the comparison results. 

\subsection{Scalability}
\changed{}{In this section, we describe how to obtain quantization schemes for high-resolution simulations via a scaling procedure. Specifically, we choose to first optimize for a quantization scheme using the corresponding low-resolution simulation and then apply the scheme to the high-resolution simulation. Its advantage is that we can avoid the expensive cost of high-resolution differentiable simulation in terms of both memory and computation. We demonstrate how to scale the resolution up in an example of 3D fluid simulation to verify the effectiveness of the scaling procedure, and the influence of the initial conditions on the quantization scheme is also evaluated.}

\changed{In practice, we use the average height of the particles as an approximation of fluid volume. to look out for the volume change.}{We use the average height as the evaluation function in this experiment since it is approximately proportional to the fluid volume at equilibrium.}  For each particle, we store the position ($\mathbf{p}$), velocity ($\mathbf{v})$, the volume ratio ($J$), and the local affine velocity field ($\mathbf{C}$). These quantities add up to a total memory consumption of $48$ bytes. Starting from $0.2$M particles, we show that the unmodified quantization scheme can deliver physically-plausible results with $0.5$M, $1$M, and $5$M particles with the configurations in Table~\ref{tab:scaling-config}. While, with the help of the bisection gradient computation algorithm, we can compute gradients of a simulation with many time steps, it is still costly to directly compute the gradients for 5M or even more particles. Therefore, we use the gradients computed with 0.2M particles. \changed{}{The ranges are obtained via running the simulation at the highest resolution in Table~\ref{tab:scaling-config}. Finally, we set the ranges of the fixed numbers to be twice the value of the recorded maximal ranges.}

\begin{figure}
  \includegraphics[width=\linewidth]{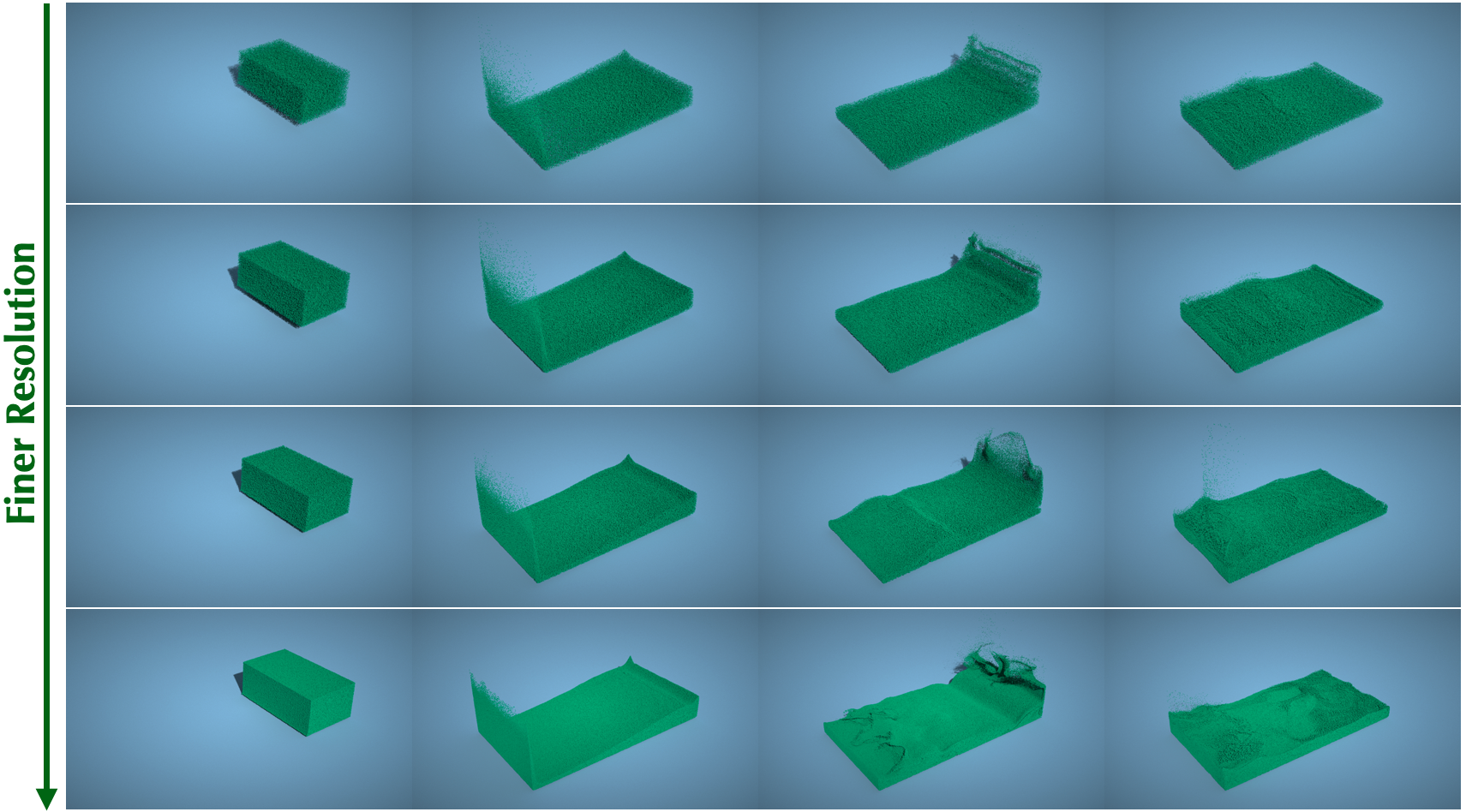}
  \caption{Visual comparisons of four simulations by gradually refining the resolution from top to bottom. All the four simulations share the same quantization scheme obtained with the first configuration in Table~\ref{tab:scaling-config}. The visual quality does not show significant degradation in the four different levels of resolution.}
  \label{fig:scaling}
\end{figure}

\begin{table}
\centering
\caption{Scaling configurations and the evaluations functions of the simulation. The reference evaluation function value is obtained by the \floatpoint{64} type. The quantized evaluation function is obtained by applying the quantization scheme derived using the first configuration. The quantization scheme can be found in the supplemental materials.}
    \resizebox{\linewidth}{!}{%
    \begin{tabular}{c|c|c|c|c|c}
    \toprule
    \textbf{Particles} & \textbf{$\Delta t$} & \textbf{Grid}  & \textbf{Steps} & \ \textbf{\begin{tabular}[c]{@{}c@{}}Reference \\ Evaluation Function \end{tabular}} & \textbf{\begin{tabular}[c]{@{}c@{}}Quantized \\ Evaluation Function \end{tabular}} \\ \midrule \midrule
    \textbf{200,000}    & $2\times 10^{-4}$  & $64^3$ & 16384 & $1.9910 \times 10^{-2}   $   & $1.9908 \pm 0.0004 \times 10^{-2}$ \\ \hline
    \textbf{500,000}    & $2\times 10^{-4}$  & $64^3$ & 16384 & $2.0130 \times 10^{-2}   $   & $2.0543 \pm 0.0019 \times 10^{-2}$ \\ \hline
    \textbf{1,000,000}    & $1\times 10^{-4}$  & $128^3$ & 32768 & $2.6773 \times 10^{-2} $  & $2.6776 \pm 0.0010\times 10^{-2}$\\ \hline
    \textbf{5,000,000}    & $5\times 10^{-5}$  & $256^3$ & 65536 & $2.8241 \times 10^{-2}$   & $3.2300 \pm 0.0015 \times 10^{-2}$\\\bottomrule
    \end{tabular}
}%
\label{tab:scaling-config}
\end{table}

\paragraph{Scaling up resolution.}
\changed{After getting both gradients and ranges ready, our system automatically generates a quantization scheme according to user-specified control parameters.}{}In this experiment, we use the \emph{memory-bounded quantization} to obtain a scheme with a memory compression rate of $50$\%.
We directly apply the scheme to the simulator with the four configurations in Table~\ref{tab:scaling-config} with increased particle densities and keep a constant initial volume in the scaling procedure. To exploit the advantages of additional particles, we refine the background grids synchronously, which leads to a smaller time step for the sake of numerical stability. We also increase the total number of steps to match the overall duration. In the end, with all these modifications on numerous parameters, we still arrive at a large-scale physically-plausible simulation, as shown in Fig.~\ref{fig:scaling} and the supplemental material.


\begin{figure}
  \includegraphics[width=\linewidth]{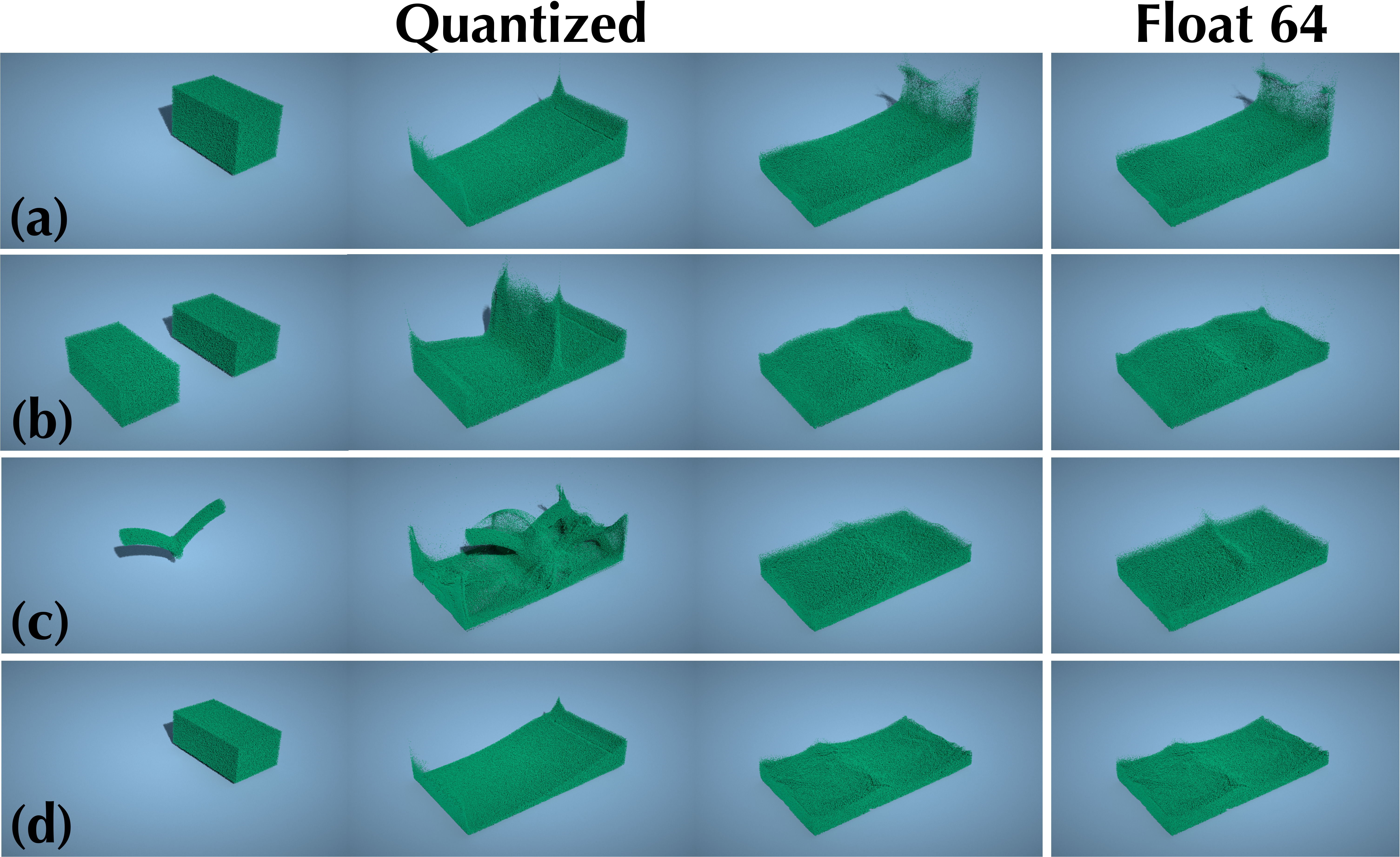}
  \caption{\changed{}{Simulation results under different initial conditions and parameters.
  The left three columns are the results of three time steps using the same quantization scheme as the scaling experiment in Fig.~\ref{fig:scaling}, and the last column shows the corresponding results of a \floatpoint{64} reference for the last time step. From top to bottom: We change the initial position and volume in rows (a) and (b); The result with an initial velocity of 2.0 is shown in row (c); The last row (d) shows the result using $4\times$ larger pressure coefficient in the fluid state equation.}}
  \label{fig:init_condition}
\end{figure}

\changed{}{
\paragraph{Generalization to different initial conditions.}
To evaluate our framework with different initial conditions, we apply the same scheme obtained in the above scaling experiment to four different initial simulation settings and show the results in Fig.~\ref{fig:init_condition}.
We use 1M particles for the four experiments and only adjust the initial position volume and velocity. We also increase the fluid pressure coefficient by a factor of $4$ to test the simulation stability (see the bottom row of Fig.~\ref{fig:init_condition}).
Our scheme can produce visual results very similar to the \floatpoint{64} reference. However, if we further decrease the compression rate from 50\% to 30\%, artifacts will appear during the simulation. For the failure case, please refer to section 2.5 in our supplementary material.
}


\subsection{Large-scale Simulations}
We show three large-scale simulation results using our method, including a 3D Eulerian smoke simulation, a 3D elastic simulation, and a fluid simulation based on MLS-MPM.
The configuration details are listed in Table \ref{tab:large_scale} and the rendered results are shown in Fig.~\ref{fig:large_scale_result}.
Their quantization schemes can be found in our supplemental materials.

\newcommand{\largescalefigurewidth}{0.196}
\begin{figure*}
  \centering

  \includegraphics[width=\largescalefigurewidth\linewidth]{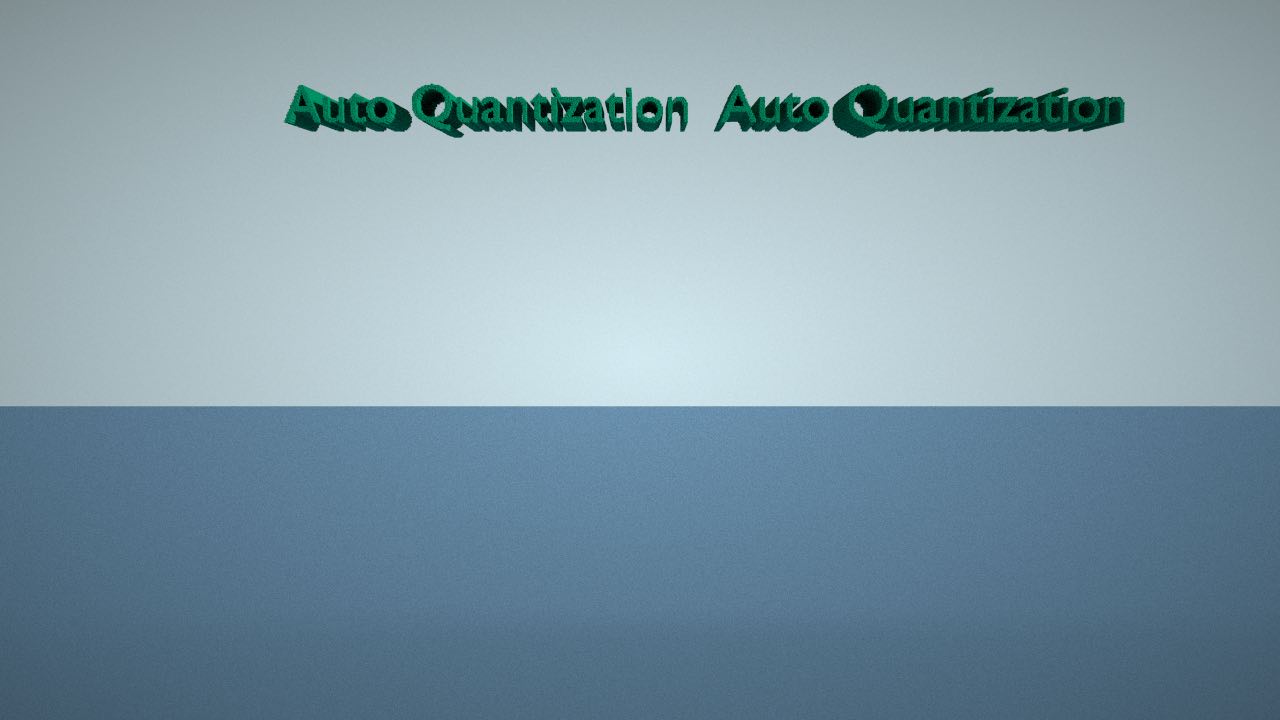}
  \includegraphics[width=\largescalefigurewidth\linewidth]{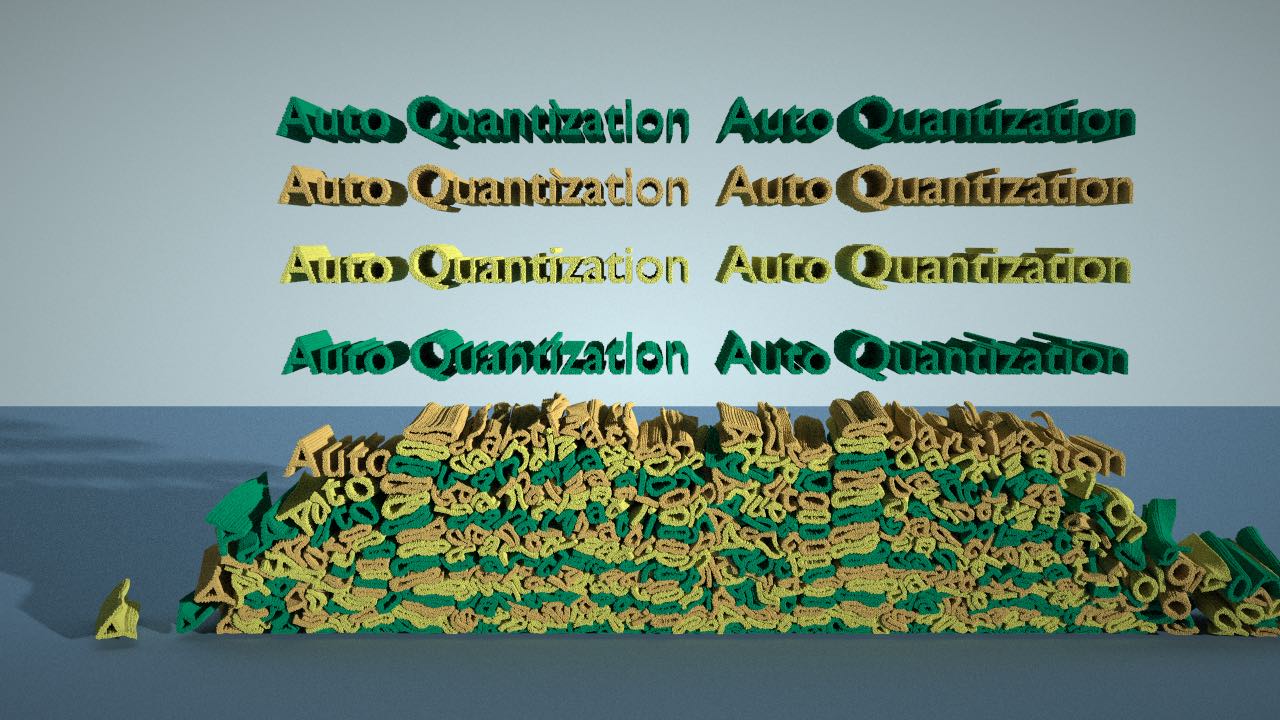}
  \includegraphics[width=\largescalefigurewidth\linewidth]{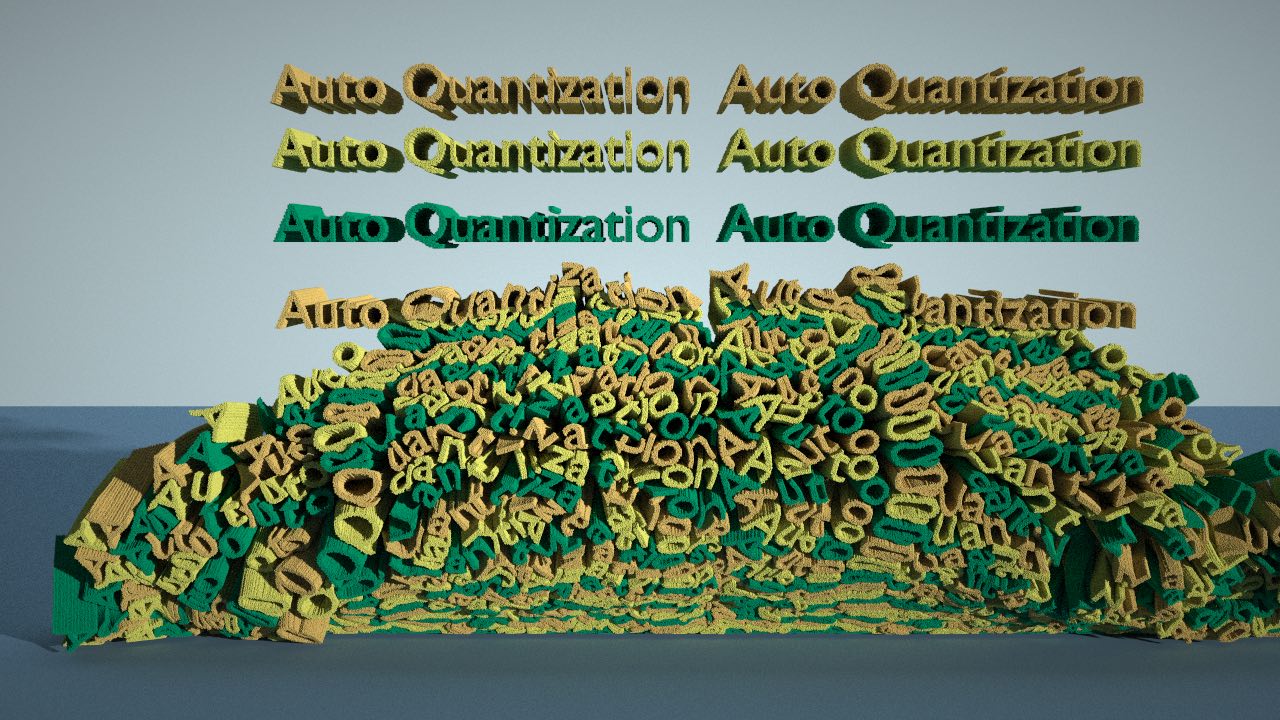}
  \includegraphics[width=\largescalefigurewidth\linewidth]{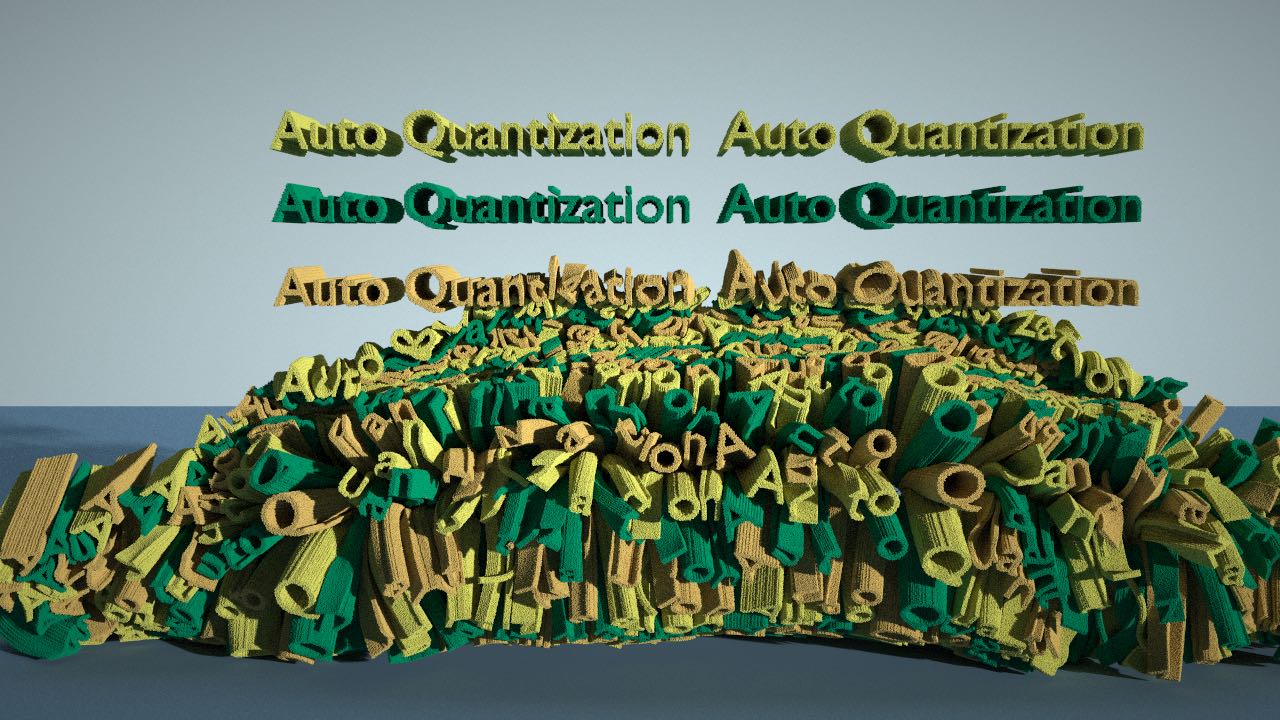}
  \includegraphics[width=\largescalefigurewidth\linewidth]{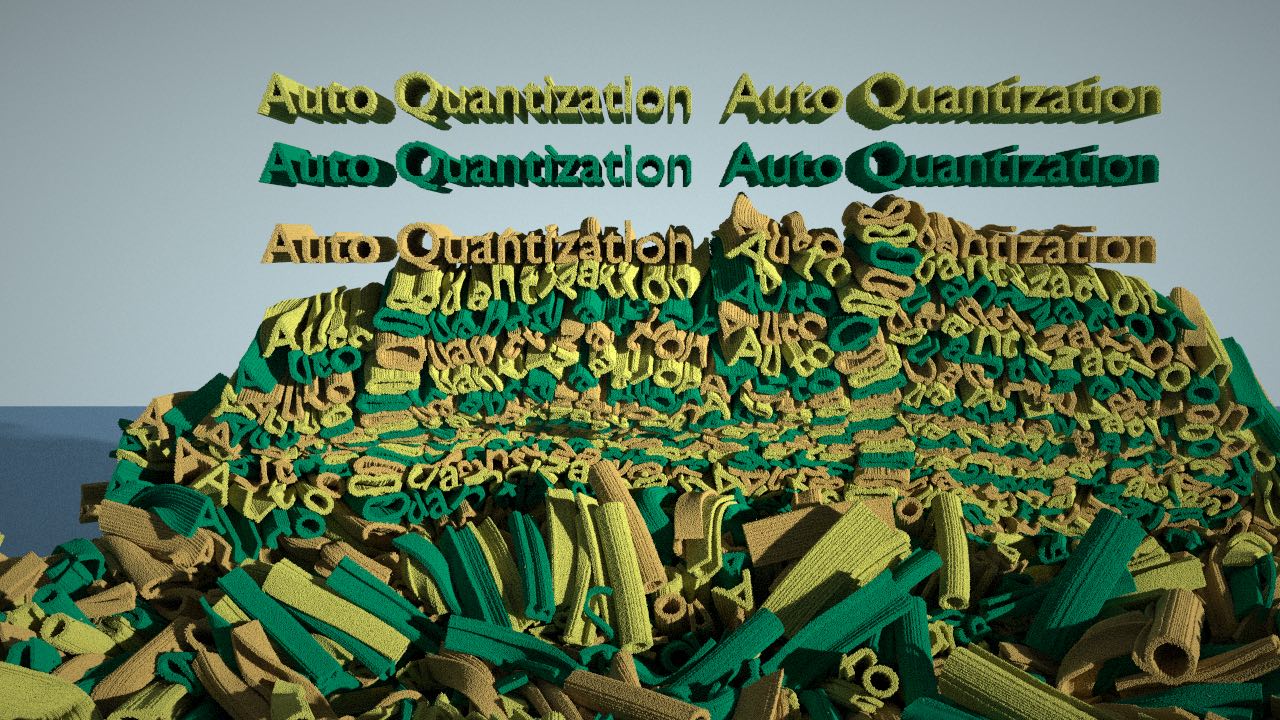}

  \includegraphics[width=\largescalefigurewidth\linewidth]{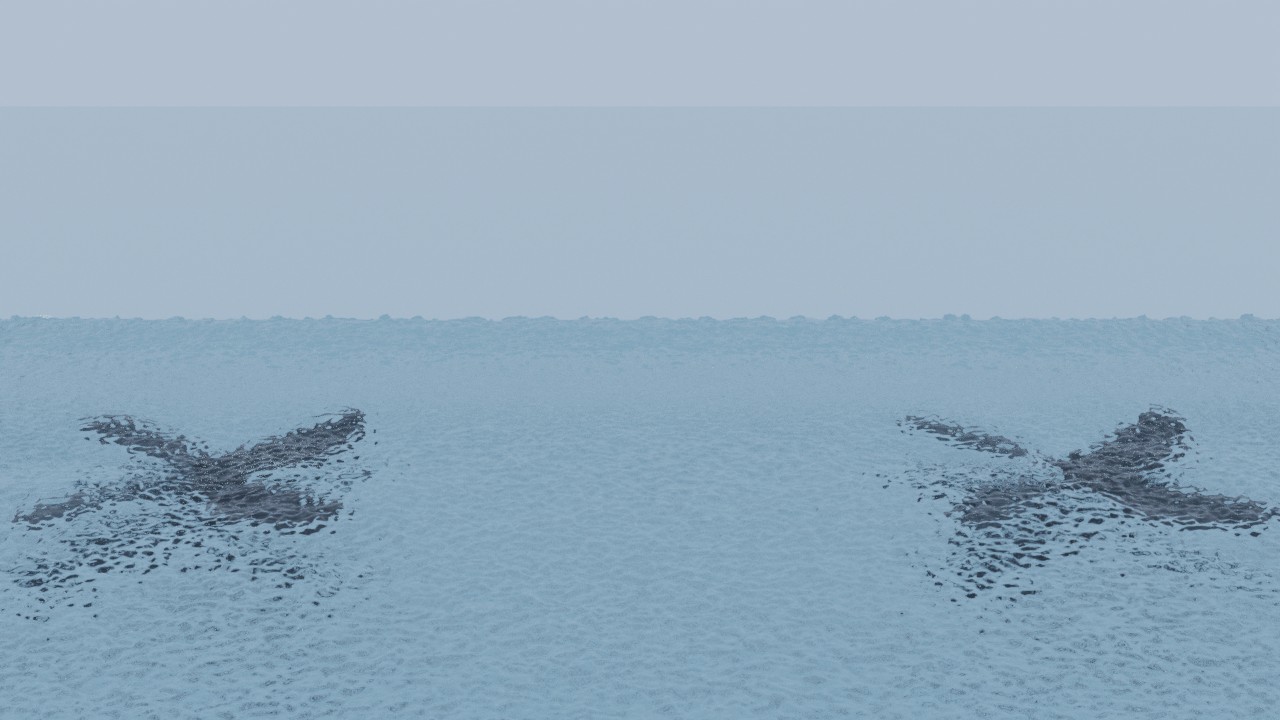}
  \includegraphics[width=\largescalefigurewidth\linewidth]{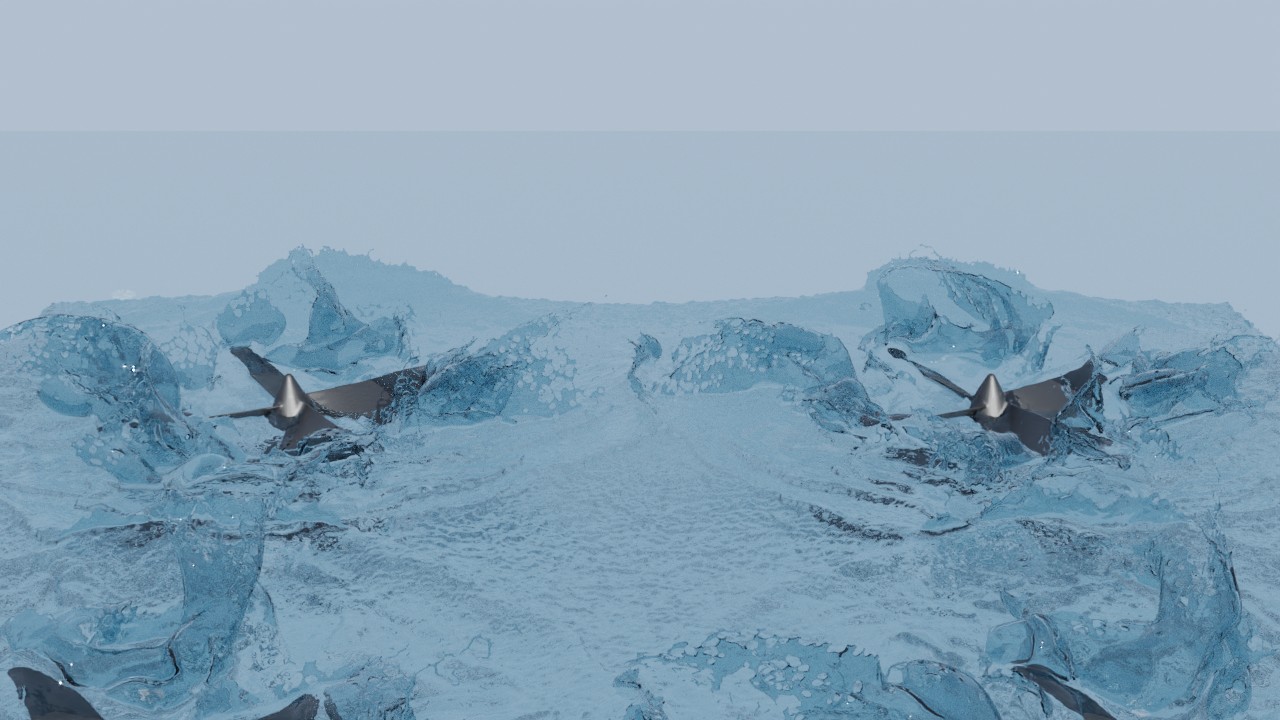}
  \includegraphics[width=\largescalefigurewidth\linewidth]{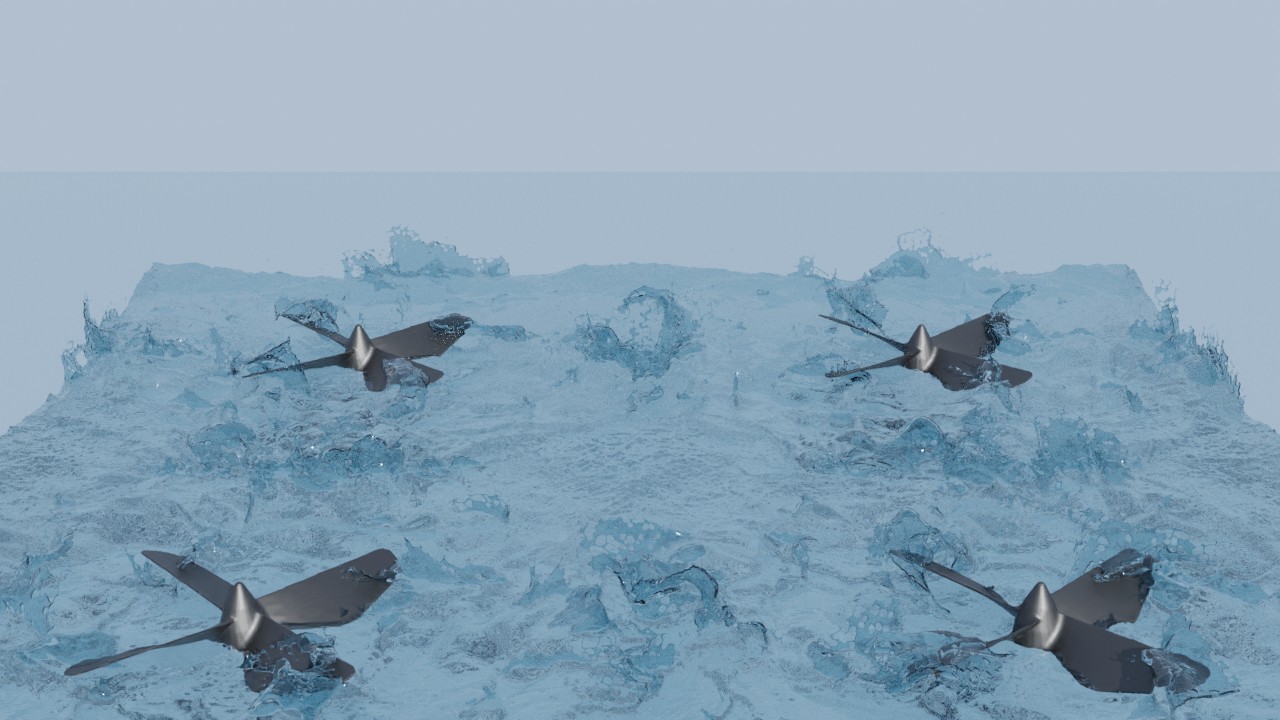}
  \includegraphics[width=\largescalefigurewidth\linewidth]{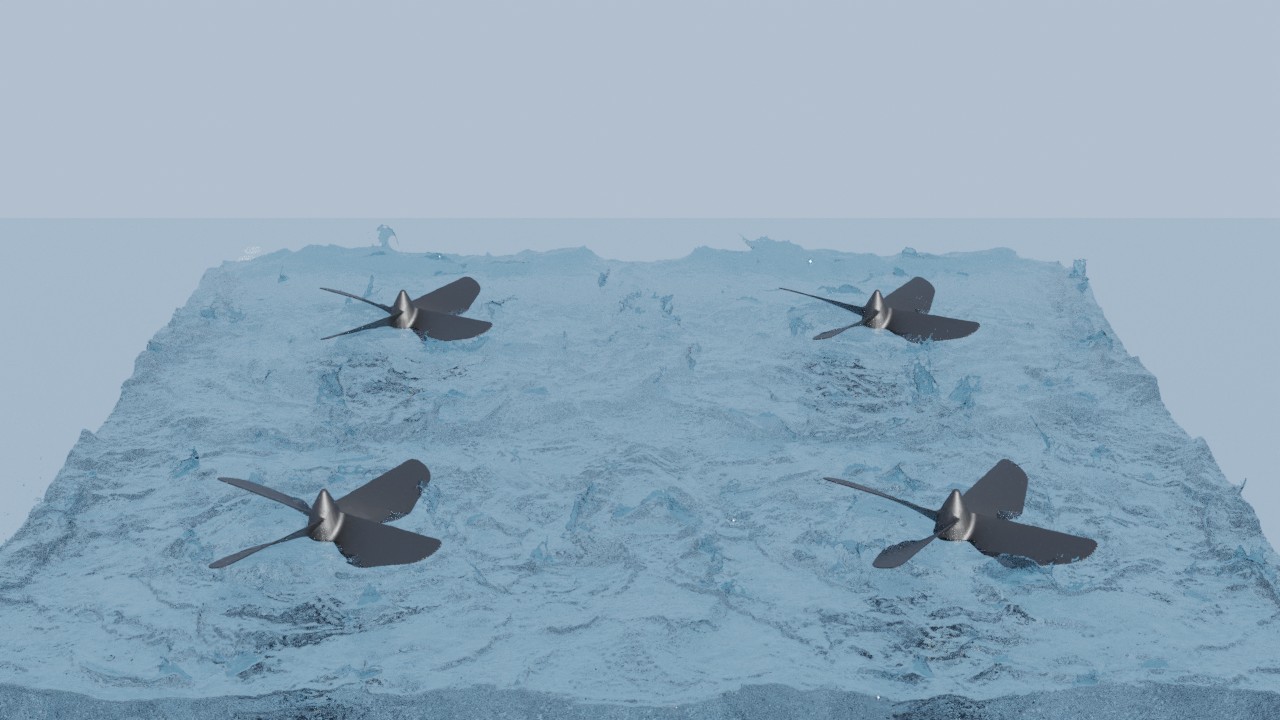}
  \includegraphics[width=\largescalefigurewidth\linewidth]{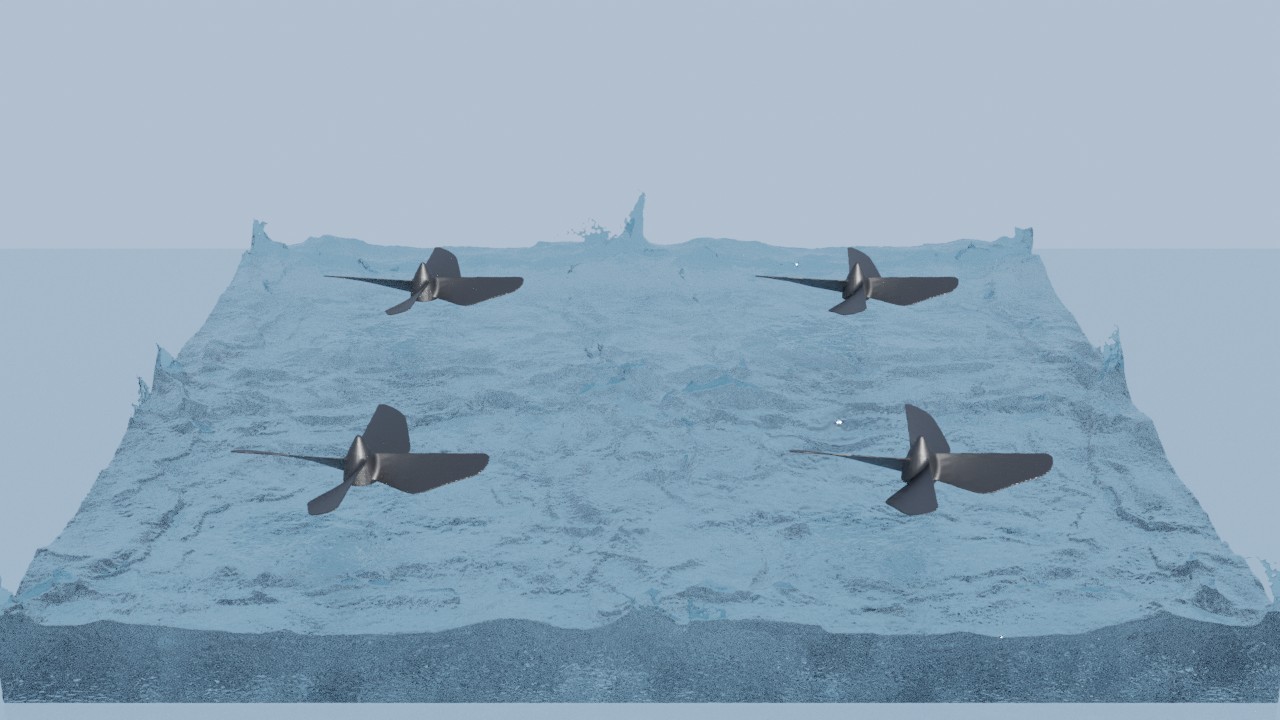}

  \includegraphics[width=\largescalefigurewidth\linewidth]{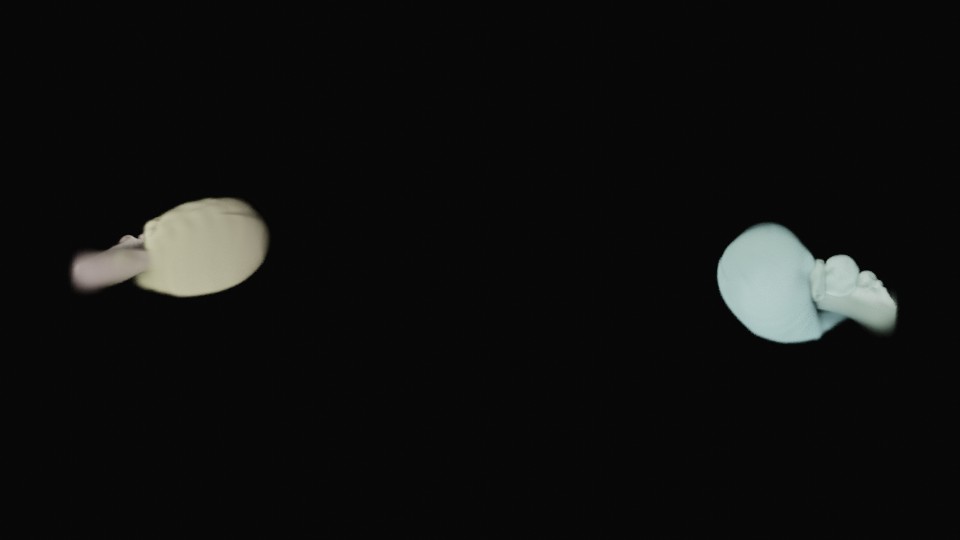}
  \includegraphics[width=\largescalefigurewidth\linewidth]{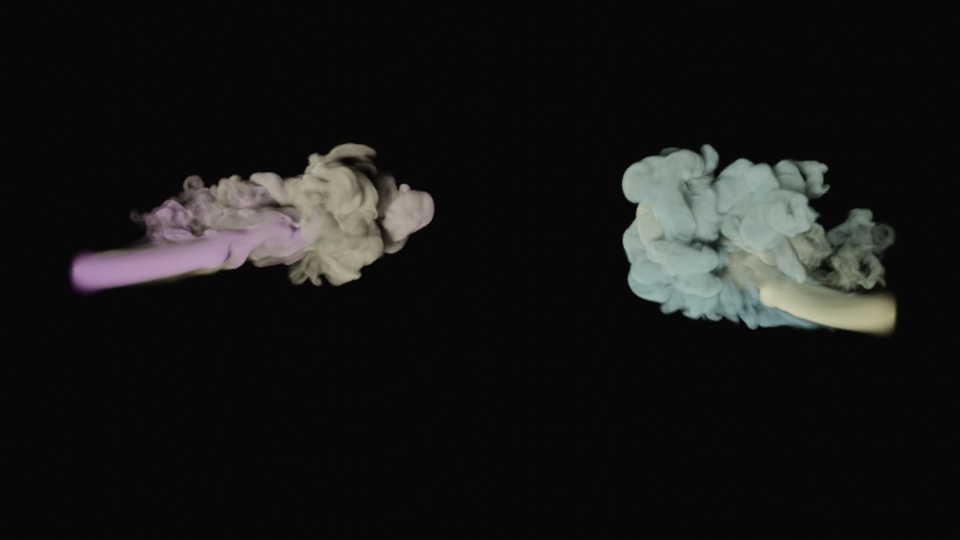}
  \includegraphics[width=\largescalefigurewidth\linewidth]{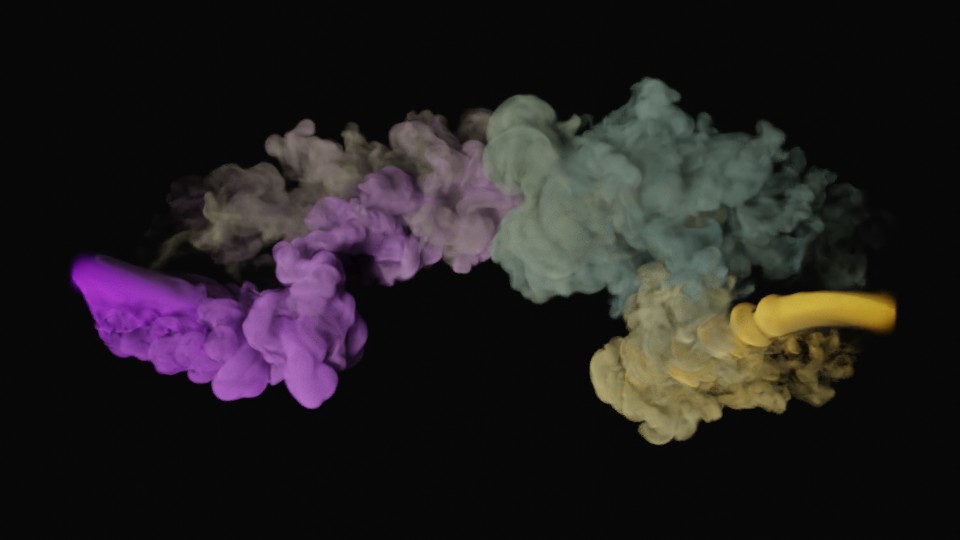}
  \includegraphics[width=\largescalefigurewidth\linewidth]{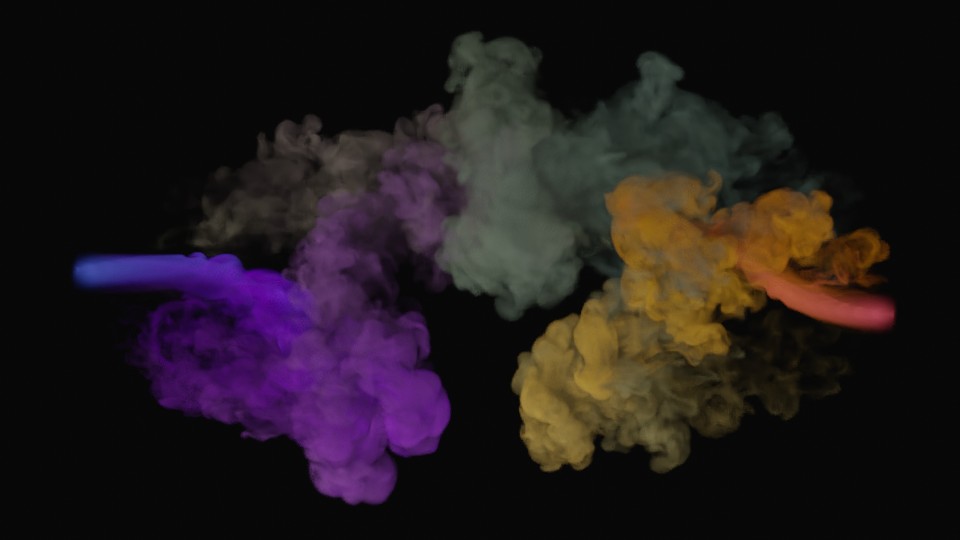}
  \includegraphics[width=\largescalefigurewidth\linewidth]{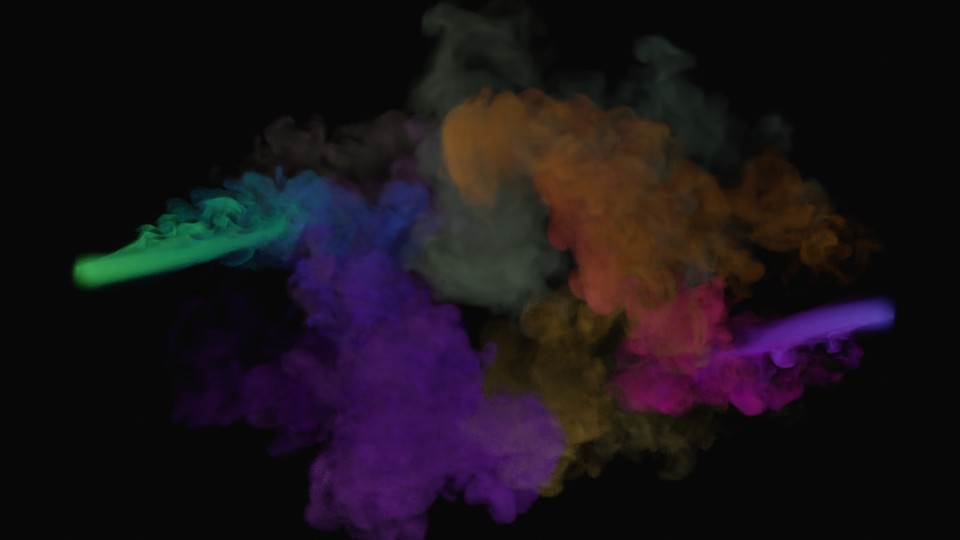}

  \caption{Snapshots of large-scale simulation results using our automatic quantized schemes.}
  \label{fig:large_scale_result}
\end{figure*}

\begin{table*}
\centering
\caption{The configuration and results of three large-scale simulations. The elastic body and liquid simulations are based on the MLS-MPM algorithms and the smoke simulation is using an advection-reflection solver.}
\resizebox{\linewidth}{!}{%
\begin{tabular}{c|cc|cc|cc|cc|c|c|c|c}
\toprule
\multirow{2}{*}{\textbf{Demo}} & \multicolumn{2}{c|}{\textbf{$\Delta t$}}                       & \multicolumn{2}{c|}{\textbf{Particles}}                & \multicolumn{2}{c|}{\textbf{Grids}}                                   & \multicolumn{2}{c|}{\textbf{Frames}}                   & \multirow{2}{*}{\textbf{\begin{tabular}[c]{@{}c@{}}Steps/\\ Frame\end{tabular}}} & \multirow{2}{*}{\textbf{\begin{tabular}[c]{@{}c@{}}Active \\ Voxels\end{tabular}}} & \multirow{2}{*}{\textbf{\begin{tabular}[c]{@{}c@{}}
Memory\\ Compression\end{tabular}}} & \multirow{2}{*}{\textbf{\begin{tabular}[c]{@{}c@{}}Seconds/\\ Frame\end{tabular}}} \\ \cline{2-9}
                               & \multicolumn{1}{c|}{\textbf{Initial}} & \textbf{Final} & \multicolumn{1}{c|}{\textbf{Initial}} & \textbf{Final} & \multicolumn{1}{c|}{\textbf{Initial}}       & \textbf{Final}          & \multicolumn{1}{c|}{\textbf{Initial}} & \textbf{Final} &                                                                                   &                                                                                    &                                                                                       &                                                                                    \\ \midrule
\textbf{Elastic body}          & \multicolumn{1}{c|}{$2\times 10^{-4}$ s}             & $7.5 \times 10^{-5} $  s       & \multicolumn{1}{c|}{1,000,000}        & 295,280,208    & \multicolumn{1}{c|}{$256^3$}                    & $1024^3$  & \multicolumn{1}{c|}{128}              & 320            & 128                                                                               & -                                                                                  & $2.01\times$                                                                                 & 63.5 s                                                                             \\ \hline
\textbf{Fluid (liquid)}        & \multicolumn{1}{c|}{$4\times 10^{-4}$ s}             & $1\times 10^{-4} s$           & \multicolumn{1}{c|}{1,000,000}        & 400,000,000    & \multicolumn{1}{c|}{$128^3$} & $256^3$  & \multicolumn{1}{c|}{128}              & 256            & 128                                                                               & -                                                                                  & $2.02\times$                                                                                 & 139.3 s                                                                            \\ \hline
\textbf{Fluid (smoke)}                 & \multicolumn{1}{c|}{0.01 s}             & 0.01 s           & \multicolumn{1}{c|}{-}                & -              & \multicolumn{1}{c|}{$128^3$} & $1024^3$ & \multicolumn{1}{c|}{256}              & 300            & 1                                                                                 & 228,982,784                                                                        & $1.93\times$                                                                                 & 49.3 s                                                                             \\ \bottomrule
\end{tabular}
}
\label{tab:large_scale}
\end{table*}

\paragraph{Eulerian smoke simulation}
We develop a large-scale advection-reflection solver with the same algorithm described in Sec.~\ref{sec:error_bound}.
We start our workflow by computing the gradients in the \floatpoint{64} implementation.
Then we apply the resulting scheme with a $1.93\times$ compression and eventually get a large-scale smoke simulation with over 228M voxels activated in total.

\paragraph{MLS-MPM simulations}
We develop two large-scale MLS-MPM solvers featuring simulation of fluid and elastic body, respectively.
Gradients are computed with 100M particles under $128^3$ and $256^3$ background grids, respectively.
The sum of the final kinetic energy and gravitational potential energy is used as the evaluation function for both simulations.
When scaling to relatively high resolution, we observe that the ranges of some attributes such as the local affine velocity $\mathbf{C}$ become much broader.
To prevent the simulator from overflowing, we increase the range by $24$ times for $\mathbf{C}$ in the fluid simulation.
Another way to determine the range of fixed-point numbers is evaluating the simulation at a relatively high resolution in full precision.
We perform an elastic body simulation at a resolution of $125$M particles and $1024^3$ background grids to obtain the ranges and scale them by a factor of two.
In the fluid simulation demo, we clamp the gradients during the gradients back-propagation to avoid outliers.
Finally, we are able to run a fluid simulation example with $400$M particles and an elastic body simulator with $295$M particles.


\section{Conclusions}

We have developed a novel automatic quantization framework to determine quantization schemes, which circumvents manual trial and error and significantly improves efficiency.
\changed{
With accounts for precision in our model, our quantization scheme is of high quality, which is confirmed in our experiments.
We hope our work will set up a reliable reference for the users who seek to use quantized types for physics based simulation.}{Users can obtain a feasible quantization scheme by simply setting a target memory compression rate or an acceptable error bound. 
Experimental results show that our method can generate quantization schemes that can balance the memory consumption and computation errors according to users' preferences, which achieves up to $2.5\times$ memory compression rate without noticeable artifacts.
} 

\paragraph{Limitations.}
In our workflow, \changed{before achieving the goal of high-resolution simulation, the scaling process is still a stumbling rock with intensive human involvement.
Currently, scaling the simulation modifies a chain of parameters and its automation is still an unsolved issue.}{the quantization scheme for a high-resolution simulation is obtained via its low-resolution version, which does not take the difference between the low-resolution result and the high-resolution counterpart into account.}
Although the procedure of scaling up the resolution achieves reasonable results in our experiments, there is no theoretical guarantee that it can work all the time.

\paragraph{Future work.}
We currently prefer efficiency over optimality and accept the slim possibility of failure to meet the error constraint. However, it is desirable to investigate how to speed up iterative optimization to enforce the error control more robustly. Moreover, we have only developed and tested our system on fixed-point representation, leaving the support of floating-point representation as a possible enhancement in the future.

\begin{acks}
We thank anonymous reviewers for their constructive comments. Weiwei Xu is partially supported by NSFC grant (No. 61732016) and Kuaishou Research Collaboration Initiative.
Yin Yang is partially supported by National Science Foundation grant (No. 2011471, 2016414).
This paper is supported by Information Technology Center and State Key Lab of CAD\&CG, Zhejiang University.
\end{acks}

\bibliographystyle{ACM-Reference-Format}
\bibliography{ref}

\appendix

\end{document}